\documentclass[extra,onecolumn]{gji}

\usepackage[fleqn]{amsmath}
\usepackage{amsfonts,amssymb}
\usepackage{timet,epsfig,wrapfig,psfrag}
\usepackage{natbib}
\usepackage{symbol}
\usepackage{xcolor}
\usepackage[normalem]{ulem}
\usepackage{booktabs}
\usepackage{multirow} 
\usepackage{pdfcolmk} 
\usepackage{braket}
\usepackage{listings}
\usepackage{dsfont}
\usepackage{placeins}
\usepackage{subcaption}

\usepackage{hyperref}
\hypersetup{
    colorlinks=true,
    linkcolor=blue,
    filecolor=magenta,      
    urlcolor=cyan,
}

\definecolor{codegray}{gray}{0.9}

\begin{document}

\bibliographystyle{chicago}


\title[1-D elastic wave propagation on quantum computers]
{A quantum computing concept for 1-D elastic wave simulation with exponential speedup}

\author[Malte Schade et al.]{
\parbox{\linewidth}{
Malte Schade,$^{1,\dagger}$
Cyrill B\"osch,$^{1,\dagger}$
V\'aclav Hapla,$^1$
Andreas Fichtner$^1$} \\
$^1$ Department of Earth and Planetary Sciences, ETH Zurich, Sonneggstrasse 5, \\ 8092 Zurich, Switzerland. Email: \href{mailto:cyrillboesch@gmail.com}{cyrillboesch@gmail.com} \\
$^\dagger$ These authors contributed equally.}
\date{07.05.2024}
\maketitle



\begin{summary}
Quantum computing has attracted considerable attention in recent years because it promises speedups that conventional supercomputers cannot offer, at least for some applications. Though existing quantum computers are, in most cases, still too small to solve significant problems, their future impact on domain sciences is already being explored now. Within this context, we present a quantum computing concept for 1-D elastic wave propagation in heterogeneous media with two components: a theoretical formulation and an implementation on a real quantum computer. The method rests on a finite-difference approximation, followed by a sparsity-preserving transformation of the discrete elastic wave equation to a Schr\"{o}dinger equation, which can be simulated directly on a gate-based quantum computer. An implementation on an error-free quantum simulator verifies our approach and forms the basis of numerical experiments with small problems on the real quantum computer IBM Brisbane. The latter produce simulation results that qualitatively agree with the error-free version but are contaminated by quantum decoherence and noise effects. Complementing the discrete transformation to the Schr\"{o}dinger equation by a continuous version allows the replacement of finite differences by other spatial discretization schemes, such as the spectral-element method. Anticipating the emergence of error-corrected quantum chips, we analyze the computational complexity of the best quantum simulation algorithms for future QCs. This analysis suggests that our quantum computing approach may lead to wave field simulations that run exponentially faster than simulations on classical computers. 
\end{summary}


\section{Introduction}\label{S:Introduction}

The progress of seismological research has always been closely linked to advances in computer technology.
As early as 1968, the then-novel IBM 360/65 enabled the random sampling of more than 200,000 seismic models per hour,
leading to one of the first Monte Carlo inversions for whole-Earth structure \citep{Press_1968}.
The Connection Machine CM-5 was the workhorse for large-scale finite-difference wave field simulations \citep{Igel_1995} and early 2-D acoustic full-waveform inversion \citep{Igel_1996}, which could be extended to massive 3-D elastic inversions when GPUs became available \citep[e.g.,][]{Bozdag_2016}. Novel chip designs that integrate CPUs and GPUs substantially reduce code complexity,
thereby facilitating the efficient development and execution of seismic applications on personal computers \citep{gebraad_seamless}.
The list of prominent examples is endless.

Emerging quantum computers (QCs) are opening a new chapter in scientific computing, even though available QCs are still relatively small and inefficient compared to the latest supercomputers that are based on large numbers of CPUs and GPUs. The current stage of the QC development is called \emph{Noisy Intermediate Scale Quantum} \citep[NISQ; coined by][]{preskill2018quantum}, since their efficiency is mainly hampered by elevated noise \citep[e.g.,][]{knill2005quantum,nielsen2010quantum,preskill2018quantum}. Recent advances in quantum error correction \citep[e.g.,][]{sivak2023real, kim2023evidence,bravyi2023high} and the design of larger, more modular QCs \citep[e.g.,][]{gibney2019hello, gambetta2020ibm, sevilla2020forecasting} are promising developments, suggesting the feasibility of quantum computation for numerous algorithmic challenges within the next decade. Presently, QCs are already available in cloud environments, like IBM Quantum, and can be accessed freely for experimental use. This accessibility facilitates the testing and validation of novel quantum algorithms.

QCs harness quantum phenomena such as entanglement and superposition for problem-solving algorithms, thereby promising up to exponential reductions in both runtime and memory complexity compared to some classical algorithms \citep[e.g.,][]{nielsen2010quantum,benenti2004principles}.
As a result, quantum computation could make it feasible to tackle classes and sizes of problems considered intractable on traditional computing hardware \citep[e.g.,][]{shor1994algorithms,harrow2009quantum,montanaro2016quantum}.
As of today seismic imaging and related uncertainty quantification problems continue to be primarily limited by computational resources. It is therefore natural and timely to ask whether the potential of QCs can be harnessed for such inverse problems (\cite{moradi2018quantum,moradi1019quantum}). Recent studies have used adiabatic quantum computers (\cite{o2018approach, golden2021pre, souza2022application, dukalski2023quantum}) and variational quantum algorithms (\cite{trahan2023variational}) for solving geoscience-related problems (see Section \ref{ss: relation to other algs} for more details). In all studies, the quantum resources are used to solve linearised inverse problems. Here, we intend to take a first step toward non-linear Full Wave Form Inversion (FWI) (\cite{Virieux_2009, fichtner2010full, Liu_2012}) by tackling the forward problem of solving the elastic wave equation, using a different algorithmic approach, known as Hamiltonian simulation (HS). HS has been invented to simulate quantum systems and is among the most promising applications of QCs with the potential for exponential speedup over classical computers (\cite{low2019hamiltonian}). It has recently been demonstrated in a handful of studies that classical wave problems can be mapped to a Schrödinger equation so that HS can be harnessed to simulate Maxwell's equations (\cite{jin2022quantum,jin2023quantum}), homogeneous acoustic waves, and mass-spring models  (\cite{suau2021practical, costa2019quantum, babbush2023exponential}). This study exploits HS to solve the 1-D elastic wave equation in heterogeneous media. Additionally, we perform initial experiments on actual QCs.

More specifically, we extend numerical wave propagation via HS \citep{costa2019quantum, suau2021practical} to heterogeneous media, a necessary step toward FWI on QCs. For this, we employ an analytical Cholesky decomposition, which facilitates the transformation of the elastic wave equation with heterogeneous medium properties to a Schrödinger equation. We proceed with the implementation of our method on the QC IBM Brisbane, in order to solve a small-scale problem and explain current challenges. These include systematic errors caused by noise and decoherence rates, which are characteristic for current NISQ QCs. Anticipating future developments, we finally delve into the potential scalability of an optimal HS algorithm on future quantum hardware equipped with error correction.
By recognizing our problem as a $d$-sparse HS (\cite{low2019hamiltonian,babbush2023exponential}), we establish that the algorithm is capable of providing exponential speedup compared to computations on classical computers.

To conclude the study, we briefly elaborate on perspectives to resolve the read-out problem, arising from the probabilistic nature of quantum mechanics. We further derive a continuous version of the classical-to-quantum map and discuss how this paves the way for employing other discretization methods. Finally, we discuss the extensions and limitations of our method and how it relates to other quantum algorithms.

\section{A brief summary of quantum computing}\label{S:Quantum Computing}

To set the stage, we begin with a condensed review of the tensor product formalism and the quantum circuit model that should be accessible to a geophysical audience.
This framework is inherently dictated by the postulates of quantum mechanics
and forms the fundamental programming model of quantum computing.
A more general formalism of density matrices, as well as more in-depth introductions, can be found in the literature (\cite{nielsen2010quantum, watrous_2018, quantum_algorithms_beginners_2022, benenti2019principles}).

\subsection{Quantum particles and state space}

Consider a single quantum particle characterized by its quantized properties, such as the atomic energy level. This particle can occupy $N$ different orthogonal basis states, which span the state space of the particle. To the basis states, we assign labels $\Sigma = \{0, \dots, N-1\}$, collectively referred to as the label set of size $|\Sigma| = N$. A particle with $N$ basis states is called an $N$-level particle. The current state of the observable is represented by the state vector $\ket{\psi} = [\alpha_0, \dots, \alpha_{N-1}]^T \in \mathbb{C}^N$, where $\alpha_k \in \mathbb{C}$ is the probability amplitude corresponding to the $k$-th basis state $\ket{k}$, written in the bra-ket or Dirac notation that is ubiquitous in quantum mechanics. It encapsulates all information about the individual particles and their number and concisely represents exponentially large state vectors. The ket $\ket{a}$ is a column vector, whereas the bra $\bra{b}$ is a row vector with complex conjugate transpose $\bra{b}^\dagger = \ket{b}$. Bra and ket form the bra-ket $\braket{b|a}$, which is the scalar product $\ket{b}^\dagger \cdot \ket{a}$. The main difference, compared to the common vector notation in linear algebra, is that the content of the bra/ket does not represent individual vector entries but some convenient collective label. The state vector $\ket{\psi}$ can be expressed in terms of the basis states $\ket{k}$ with $k=0, ..., N-1$ as the linear combination $\ket{\psi} = [\alpha_0, \dots, \alpha_{N-1}]^T = \alpha_0\ket{0} + \alpha_0\ket{1} + \dots + \alpha_{N-1}\ket{N-1} \in \mathbb{C}^N$, which is called a superposition state.

The state vector of a system composed of $n$ particles is given by
\begin{equation}\label{eq:general state}
\ket{\psi} = \ket{\psi_0} \otimes \ket{\psi_1} \dots \otimes \ket{\psi_{n-1}} = \bigotimes_{j=0}^{n-1} \ket{\psi_j} \in \mathbb{C}^{N}, \qquad
N = N_0 N_1 \dots N_{n-1} = \prod_{j=0}^{n-1} N_j\,,
\end{equation}
where $\ket{\psi_j}\in \mathbb{C}^{N_j}$ is the state of the $j^\text{th}$ particle, and $N_j$ is the number of states that it can occupy. We reuse the symbols $N$ and $\psi$, as a single-particle system is a special case of a multi-particle system. The symbol $\otimes$ denotes the Kronecker or tensor product. For two arbitrary states $\mathbf{u}$ and $\mathbf{v}$, it is defined as
\begin{equation}\label{eq:tensor product 1}
\mathbf{u} \otimes \mathbf{v} = \begin{bmatrix}
    u_0 \mathbf{v}\\
    \vdots\\
    u_{N_\mathbf{u} - 1} \mathbf{v}\\
\end{bmatrix}
\in \mathbb{C}^{N_\mathbf{u} N_\mathbf{v}}\,,
\end{equation}
where $N_\mathbf{u}$ and $N_\mathbf{v}$ are the lengths of the vectors $\mathbf{u}$ and $\mathbf{v}$, respectively. Now, we introduce the kets with integer strings. If we assume basis vectors $\ket{j}$ and $\ket{k}$ having label sets $\Sigma$ and $\Gamma$, respectively, we have
\begin{equation}\label{eq:tensor product 2}
\ket{j} \otimes \ket{k} = \ket{jk}\,,
\end{equation}
where $jk$ is a string representing $(j,k) \in \Sigma \times \Gamma$ with the index $I = j \cdot |\Gamma| + k$ in the alphabetical ordering of $\Sigma \times \Gamma$, and $\ket{jk}$ is a vector with $1$ at the $I^\text{th}$ position and zeros otherwise. The tensor product of two or more superposition states can be deduced from \eqref{eq:tensor product 2} using multi-linearity.

Quantum computing typically uses particles with two basis states,
which is a natural choice because it is the minimum number for a non-trivial system
and the easiest to distinguish and manipulate in a hardware implementation.
Such a particle is referred to as a qubit in analogy to the classical bit.
Just like a bit, a qubit is an abstract object that can be mapped to a concrete physical object in many different ways.
The two standard basis states $\ket{0} = [1,0]^T$ and $\ket{1} = [0,1]^T$ can be represented, for instance,
by the spin-up and spin-down of a spinor or two disjoint subsets of atomic energy levels.
A single qubit $\ket{\psi}$ can be in an arbitrary superposition of the two states,
$\ket{\psi} = \beta_0 \ket{0} + \beta_1 \ket{1} = [\beta_0,\beta_1]^T \in \mathbb{C}^2$,
in contrast to a classical bit, which can only be in either of two states, 0 or 1.
A state of an $n$-qubit system with qubits $\ket{\psi_j}$, $j=0,...,n-1$, is again given by \eqref{eq:general state}
with $N = 2^n$, because $N_j = 2$ for every qubit. We can see that the quantum state space dimension $N$ grows exponentially with the number of qubits $n$.
This is the main source of the quantum computing potential.

Any quantum state must be normalized, i.e., satisfy the normalization condition
\begin{equation}\label{eq: normalisation condition}
||\ket{\psi}|| = \sum_{k=0}^{N-1} |\alpha_k|^2 = 1.
\end{equation}
Combining normalized states with the $\otimes$ product, the resulting composite state is normalized as well.
If $\alpha_k \neq 0$ for at least two different indices $k$,
then the state $\ket{\psi}$ simultaneously holds multiple potential outcomes
$\Omega = \{\ket{k}: \alpha_k \neq 0\}$
and is called a superposition state.
In contrast to classical systems, a measurement of the quantum system causes the state $\ket{\psi}$ to collapse from the superposition into one of the possible basis states $\ket{k} \in \Omega$ randomly with probability $P(k) = |\alpha_k|^2$, which explains the term probability amplitude used for the coefficients $\alpha_k$.
The type of measurement just introduced is a full-system measurement in the standard basis, which is the simplest case;
we do not deal with more general cases, which can be found in the literature at the top of this section.
Measurements allow us to read the label of the collapsed state in the form of classical bits, forming the only interface between quantum and classical information.

\subsection{Quantum gates and circuits}

Quantum computations consist of reversible, information-preserving transformations, represented by matrices $\mathbf{U}$ that are unitary, i.e., that satisfy $\mathbf{U}^{-1} = \mathbf{U}^{\dagger}$, where $\dagger$ represents a conjugate transpose. A gate-based QC implements a set of gates, which are hardware implementations of unitary operations forming a quantum-Turing complete set. These operations can simulate an arbitrary quantum algorithm, given a sufficient time and number of qubits. The only non-unitary QC operation is the measurement.

In practical quantum programming, a somewhat larger, redundant set of quantum gates is used for more convenient programming but then automatically translated into the minimum set supported by the hardware as needed before execution.
These basic gates form the ``quantum assembly language''.
They include single-qubit gates of size $2 \times 2$ such as the identity gate $\mathbf{G}_\textrm{I}$, the Hadamard gate $\mathbf{G}_\textrm{H}$, the NOT gate $\mathbf{G}_\textrm{X}$ or rotation gates,
and multi-qubit gates such as
$\mathbf{G}_\textrm{CNOT}, \mathbf{G}_\textrm{SWAP} \in \mathbb{C}^{4 \times 4}$
or the Toffoli gate $\mathbf{G}_\textrm{CCNOT} \in \mathbb{C}^{8 \times 8}$.

Any quantum algorithm is formed by a combination of basic gates (possibly used repeatedly)
that are composed into a quantum circuit, represented by a unitary matrix $\mathbf{W}$ formed by a product of $D$ layers $\mathbf{W}_j$,
\begin{equation}\label{eq: circuit layers}
\mathbf{W} = \mathbf{W}_{D-1} \mathbf{W}_{D-2} \dots \mathbf{W}_0 = \prod_{j=D-1}^{0} \mathbf{W}_j \quad \in \mathbb{C}^{N \times N}\,,
\end{equation}
where $D$ is called the circuit depth. The layers are evaluated from right to left in time, meaning that the indices reflect the order of evaluation. Further, each layer $\mathbf{W}_j$ is formed by a tensor product of basic gates,
\begin{equation}\label{eq: layer tensor product}
\mathbf{W}_j = \mathbf{G}_{j,t_0} \otimes \mathbf{G}_{j,t_1} \otimes \dots \otimes \mathbf{G}_{j,t_i} \otimes \dots \quad \in \mathbb{C}^{N \times N} \,,
\end{equation}
where $t_i \in \{\textrm{I}, \textrm{H}, \textrm{X}, \textrm{CNOT}, ...\}$ is the type of the (basic) gate
and $\otimes$ is the extension of the Kronecker product from \eqref{eq:tensor product 1} to arbitrary matrices
$\mathbf{A} \in \mathbb{C}^{M_\mathbf{A} \times N_\mathbf{A}}$
with entries denoted as $a_{i,j}$,
and $\mathbf{B} \in \mathbb{C}^{M_\mathbf{B} \times N_\mathbf{B}}$ as
\begin{equation}
\mathbf{A} \otimes \mathbf{B} = \begin{bmatrix}
a_{0,0} \mathbf{B} & \cdots & a_{0,N_\mathbf{A}-1} \mathbf{B}\\
\vdots & \ddots & \vdots\\
a_{M_\mathbf{A}-1,0} \mathbf{B}  & \cdots & a_{M_\mathbf{A}-1,N_\mathbf{A}-1} \mathbf{B}\\
\end{bmatrix}
\in \mathbb{C}^{M_\mathbf{A} M_\mathbf{B} \times N_\mathbf{A} N_\mathbf{B}}\,.
\end{equation}
The number of factors in the tensor multi-product \eqref{eq: layer tensor product} is at most $n$
($n$ if every basic gate of the layer acts on a single qubit),
but it can vary between layers based on the basic gates used;
the only requirement is that the total size is $N$ so that \eqref{eq: circuit layers} makes sense.
The art of finding the most efficient circuit for the given task,
i.e., the representation with the lowest $n$ and $D$,
is a subject of today's research in quantum computing.

\subsection{Hamiltonian simulation}\label{S:Hamiltonian simulation}

The dynamics of any quantum system are governed by the Schr\"{o}dinger equation
\begin{equation}\label{eq: SE}
    i\hbar \partial_t \ket{\psi} = \mathbf{H}\ket{\psi}\,,
\end{equation}
where $i$ denotes the imaginary unit, and $\mathbf{H} \in \mathbb{C}^{N \times N}$ is a Hermitian linear operator, called the Hamiltonian, which controls the time evolution of the system. The solution of (\ref{eq: SE}) can symbolically be written using the matrix exponential,
\begin{equation}\label{eq: intro time evolution}
    \ket{\psi(t)} = e^{-i\mathbf{H}t}\ket{\psi(0)}\,,
\end{equation}
with the initial state $\ket{\psi(0)}$. The time evolution operator $\mathbf{U}(t) = \exp(-i\mathbf{H}t)$ is unitary. Our approach to elastic wave propagation on a QC rests on the quantum or Hamiltonian simulation (HS) approach \citep[e.g.,][]{georgescu2014quantum, nielsen2010quantum}, which amounts to the direct evaluation of \eqref{eq: intro time evolution}. HS was originally introduced to model multi-particle quantum dynamic systems using QCs. Due to the aforementioned exponential space growth, quantum dynamics modeling becomes prohibitively complex on classical computers. In contrast, by being quantum systems themselves, QCs naturally feature the same state space expansion, potentially making multi-particle simulations possible.

For initialization, HS requires the application of a unitary operation $\mathbf{V}$ that sets the desired initial state, i.e. $\ket{\psi(0)} = \mathbf{V}\ket{00..00}$. Subsequently, the unitary time evolution operator $\mathbf{U}(t)$, constructed using basic gates, is applied to $\ket{\psi(0)}$,
\begin{equation}
    \ket{\psi(t)} = \mathbf{U}(t) \ket{\psi(0)}\,.
\end{equation}
The main challenge of HS is that the matrix exponential $\mathbf{U}(t) = \exp(-i\mathbf{H}t)$ is notoriously difficult to evaluate exactly.
Hence, the application of $\mathbf{U}(t)$ is replaced by some unitary $\mathbf{U}_\textrm{A}(t)$
that evolves the initial state to the final one with the desired error $\varepsilon$,
$||\mathbf{U}(t)\ket{\psi(0)} - \mathbf{U}_\textrm{A}(t)\ket{\psi(0)}|| < \varepsilon$,
and can be implemented efficiently using basic quantum gates,
with circuit depth growing along with $1/\varepsilon$. 

Since the matrix exponential can be expressed in terms of a Taylor series,
$\mathbf{U}(t) = e^{-i\mathbf{H}t} = \sum_{k=0}^{\infty} \frac{(-i\mathbf{H}t)^k}{k!}$,
we can approximate it by truncating it to a finite Taylor sum,
$\mathbf{U}_\textrm{A}(t) = \sum_{k=0}^{m-1} \frac{(-i\mathbf{H}t)^k}{k!} \approx e^{-i\mathbf{H}t}$, with some integer $m<\infty$.
The finite sum is then further approximated by decomposing $\mathbf{H}$ into a sum of unitary matrices.
The number of terms in this sum, determined by the given $\mathbf{H}$ and desired accuracy, dictates the efficiency of this particular approach \citep{berry2015truncated}.

Other approaches can be found in the literature, with efficiency varying based on properties of the given $\mathbf{H}$.
These include methods based on the Trotter-Suzuki Decomposition
\citep[e.g.,][]{trotter_product_1959, suzuki_generalized_1976, suzuki_general_1991, hatano2005finding, dhand_stability_2014, yi_spectral_2022}
and the recent qubitization approach \citep{low2019hamiltonian}, discussed more in detail in appendix \ref{A:Sparse Hamiltonian simulation}.

It has been shown that HS can achieve up to exponential speedup compared to analogous classical algorithms \citep[e.g.,][]{berry2007efficient, childs2018toward, daley2022practical, babbush2023exponential, low2019hamiltonian}.
We will discuss this matter as well as the question of utilizing the final state in section \ref{S:Discussion}.

In addition to the simulation of quantum systems, HS may also be harnessed to solve classical equations of motion, provided that an invertible transformation to the Schr\"{o}dinger equation can be found \citep[e.g.,][]{costa2019quantum, suau2021practical, babbush2023exponential}. Such transformations exist for classical wave equations \citep{susstrunk2016classification}, which share key properties with the Schr\"{o}dinger equation. Both describe wave-like phenomena and are linear. In the following, we will describe such a transformation for the elastic wave equation with heterogeneous medium parameters.

\section{Elastic waveform propagation using HS}\label{S:Theory}

We consider a 1-D elastic wave equation in the spatial domain $x\in[0, W]$ subject to Dirichlet and Neumann boundary conditions on the right and left, respectively,
\begin{equation}
    \rho(x)\,\partial_{tt} u(x,t) = \partial_x[\mu(x)\,\partial_x]u(x,t), \quad \partial_x u(0,t) = 0, \quad u(W,t) = 0,
\end{equation}
with time $t$, the wave field $u(x,t)$, density $\rho(x)$ and elastic modulus $\mu(x)$. Furthermore, the wave field is subject to the initial conditions $u_0 = u(x,0)$ and $v_0 = \partial_t u(x,0)$. We introduce a first-order finite-difference discretization in space with $2^{n-1}$ grid points, each with position and velocity, and a grid spacing of $\Delta x = W/(2^{n-1}-1)$, thereby approximating the continuous wave field $u(x,t)$ by the discrete wave field vector $\mathbf{u}(t) = [u(x_0,t),\dots,u(x_{2^{n-1}-1},t)]^T$. In section \ref{SS:continuous}, we discuss the potential use of other discretization methods. We further define the positive definite mass matrix $\mathbf{M}$ that serves as a discrete representation of the density distribution,
\begin{equation}\label{eq: mass matrix}
   \mathbf{M}=
   \begin{bmatrix}
    \rho(x_0) & 0 & \dots & 0 \\
    0 & \rho(x_1) & \dots & 0 \\
    \vdots & \vdots & \ddots & \vdots \\
    0 & 0 & \dots & \rho(x_{2^{n-1}-1})
    \end{bmatrix}\,,
\end{equation}
and the positive definite elasticity matrix $\mathbf{E}$, which is the discrete version of $\mu(x)$,
\begin{equation}\label{eq: elastic modulus matrix}
    \mathbf{E} = \begin{bmatrix}
    \mu(x_0) & 0 & \dots & 0 \\
    0 & \mu(x_1) & \dots & 0 \\
    \vdots & \vdots & \ddots & \vdots \\
    0 & 0 & \dots & \mu(x_{2^{n-1}-1})
    \end{bmatrix}\,.
\end{equation}
To obtain the stiffness matrix $\mathbf{K}$, we use a second-order stencil with the stress given by $ \sigma_i = \mu(x_i) [u(x_{i+1})-u(x_i)]/\Delta x$
and the elastic forces by $\partial_x \sigma(x_i) = [\sigma(x_i) -\sigma(x_{i-1})]/\Delta x$ \citep[e.g.,][]{moczo2014finite}.
We define the first-order accurate forward finite-difference matrix $\mathbf{D}$ as
\begin{equation} \label{eq: FD matrix}
   \mathbf{D} = \frac{1}{\Delta x} \begin{bmatrix}
  -1 & 1 & 0 & \dots & 0 \\
  0 & -1 & 1 & \dots & 0 \\
  0 & 0 & -1 & \dots & 0 \\
  \vdots & \vdots & \vdots & \ddots & \vdots \\
  0 & 0 & 0 & \dots 0 & -1
  \end{bmatrix}.
\end{equation}
The symmetric negative-definite stiffness matrix $\mathbf{K}$ is consequently obtained as
\begin{equation} \label{eq: FD stiffness}
\mathbf{K} = -\mathbf{D}^T\mathbf{E} \mathbf{D}.
\end{equation}
This implementation naturally results in a Dirichlet and a Neumann boundary condition on the right and left, respectively, assuming two ghost nodes. In the interest of a smooth derivation, we defer the case of arbitrary boundary conditions to appendix \ref{A:General Boundary Conditions}. With these definitions, the space-discretized wave equation reads
\begin{equation}\label{eq: discretized elastic wave eq}
    \mathbf{M}\, \partial_{tt} \mathbf{u} = \mathbf{K} \mathbf{u}\,.
\end{equation}
To solve (\ref{eq: discretized elastic wave eq}) on a QC, it needs to be transformed into the Schr\"{o}dinger equation (\ref{eq: SE}).
For this, we first apply a transformation that utilizes the positive definite, diagonal square root matrix $\mathbf{M}^{1/2}$,
which produces the mass-transformed wave field $\tilde{\mathbf{u}} = \mathbf{M}^{1/2} \mathbf{u}$ and the mass-transformed stiffness matrix
$\tilde{\mathbf{K}} = \mathbf{M}^{-1/2} \mathbf{K} \mathbf{M}^{-1/2}$.
The latter inherits the property of being symmetric negative-definite from $\mathbf{K}$.
We proceed by transforming the second-order system into a first-order system
\begin{equation} \label{eq: discrete first order}
    \partial_t \begin{bmatrix}
      \tilde{\mathbf{u}} \\ 
      \tilde{\mathbf{v}}
   \end{bmatrix}= 
   \begin{bmatrix}
      \mathbf{0} & \mathbf{\mathds{1}}_{2^{n-1}\times2^{n-1}}\\
      \tilde{\mathbf{K}} & \mathbf{0} 
   \end{bmatrix}
   \begin{bmatrix}
      \tilde{\mathbf{u}} \\ 
      \tilde{\mathbf{v}}
   \end{bmatrix} = \mathbf{Q} 
    \begin{bmatrix}
      \tilde{\mathbf{u}} \\ 
      \tilde{\mathbf{v}}
   \end{bmatrix},
\end{equation}
where $\tilde{\mathbf{v}} = \partial_t \tilde{\mathbf{u}} = \mathbf{M}^{1/2} \partial_t \mathbf{u} \in \mathbb{R}^{2^{n-1}}$ is the mass-transformed velocity vector and $\mathbf{Q}$ the impedance matrix.
While (\ref{eq: discrete first order}) is already closer to the Schr\"{o}dinger equation than (\ref{eq: discretized elastic wave eq}), it remains to symmetrize $\mathbf{Q}$ and obtain the imaginary unit $i$ on the left-hand side.
For this, we multiply (\ref{eq: discrete first order}) by an arbitrary invertible transformation $\mathbf{T} \in \mathbb{R}^{2^{n}\times2^{n}}$, which results in
\begin{equation}
    \partial_t \mathbf{T} \begin{bmatrix}
      \tilde{\mathbf{u}} \\ 
      \tilde{\mathbf{v}}
   \end{bmatrix}= 
   \mathbf{T}
    \mathbf{Q}
   \mathbf{T}^{-1}
   \mathbf{T}
   \begin{bmatrix}
      \tilde{\mathbf{u}} \\ 
      \tilde{\mathbf{v}}
   \end{bmatrix}.
\end{equation}
If $\mathbf{T}$ can be constructed such that $\mathbf{H} = i\, \mathbf{T}\mathbf{Q}\mathbf{T}^{-1}$ is Hermitian,
we can employ $\mathbf{H}$ as a quantum Hamiltonian that emulates elastic wave field dynamics.
The transformed wave field vector is identified as $\ket{\phi} = \mathbf{T}[\tilde{\mathbf{u}},\tilde{\mathbf{v}}]^T$,
which is subsequently encoded in a quantum state. The final Schr\"{o}dinger equation for 1-D elastic wave propagation therefore becomes
\begin{equation}\label{eq: quantum evolution}
    i\, \partial_t \ket{\phi(t)} = \mathbf{H} \ket{\phi(t)},
\end{equation}
subject to the initial conditions
\begin{equation}\label{eq: transformed initial}
    \ket{\phi(0)} = \mathbf{T} \begin{bmatrix}
      \mathbf{M}^{1/2}\mathbf{u_0} \\ 
      \mathbf{M}^{1/2}\mathbf{v_0}
   \end{bmatrix}\,,
\end{equation}
which are typically sparse in seismic applications thanks to the spatial localization of sources. Furthermore, owing to the localized interactions of the discretized degrees of freedom, both the mass and stiffness matrices are sparse, which implies that $\mathbf{Q}$ is sparse, too. Eq. (\ref{eq: quantum evolution}) can, in principle, be solved by a QC, and it is isomorphic to the discretized elastic wave equation (\ref{eq: discretized elastic wave eq}) because $\mathbf{T}$ is invertible. The actual elastic wave field at all times can, therefore, be reconstructed via
\begin{equation}
[\tilde{\mathbf{u}}(t),\tilde{\mathbf{v}}(t)]^T = \mathbf{T}^{-1}\ket{\phi(t)}
\qquad\mbox{and}\qquad
\mathbf{u}(t) = \mathbf{M}^{-1/2} \tilde{\mathbf{u}}(t).
\end{equation}
It remains to design a practically useful transformation $\mathbf{T}$. Several transformations exist that achieve this classical-to-quantum mapping \citep[e.g.,][]{susstrunk2016classification,kane2014topological,babbush2023exponential}. In the subsequent analysis, we choose a transformation that retains sparsity in both the initial conditions (\ref{eq: transformed initial}) and the Hamiltonian $\mathbf{H}$. This preservation of sparsity is pivotal for scalable HS algorithms. We will revisit this topic in appendix \ref{A:Sparse Hamiltonian simulation}.  To construct the desired $\mathbf{T}$, we define
\begin{equation}
   \mathbf{U} = \mathbf{E}^{1/2}\mathbf{D}\mathbf{M}^{-1/2},
\end{equation}
such that 
\begin{equation}\label{eq: analytic cholesky}
    \tilde{\mathbf{K}} = -\mathbf{U}^T\mathbf{U} = -\mathbf{M}^{-1/2}\mathbf{D}^T\mathbf{E}\mathbf{D}\mathbf{M}^{-1/2},
\end{equation}
which is an analytical Cholesky decomposition, with $\mathbf{U}$ being the upper triangular matrix.
Since the negative-definite $\tilde{\mathbf{K}}$ is invertible, $\mathbf{U}$ is also invertible.
This follows from $\text{det}(\tilde{\mathbf{K}})=\text{det}(-\mathbf{U}^T\mathbf{U}) = \text{det}(-\mathbf{U}^T)\text{det}(\mathbf{U}) $.
With this, we finally define
\begin{equation}\label{eq: discrete transformation}
    \mathbf{T} = \begin{bmatrix}
      \mathbf{U} & \mathbf{0}\\
      \mathbf{0} & \mathbf{\mathds{1}}_{2^{n-1}\times2^{n-1}}
   \end{bmatrix},
\end{equation}
which leads to 
\begin{equation}
     \mathbf{H} = i\, \mathbf{T}\mathbf{Q}\mathbf{T}^{-1}
   = 
   i\, \begin{bmatrix}
      \mathbf{U} & \mathbf{0}\\
      \mathbf{0} & \mathbf{\mathds{1}}_{2^{n-1}\times2^{n-1}}
   \end{bmatrix}
   \begin{bmatrix}
      \mathbf{0} & \mathbf{\mathds{1}}_{2^{n-1}\times2^{n-1}}\\
      -\mathbf{U}^T\mathbf{U} & \mathbf{0} 
   \end{bmatrix}
   \begin{bmatrix}
      \mathbf{U}^{-1} & \mathbf{0}\\
      \mathbf{0} & \mathbf{\mathds{1}}_{2^{n-1}\times2^{n-1}}
   \end{bmatrix}
   = 
   i\,\begin{bmatrix}
      \mathbf{0} & \mathbf{U}\\
      -\mathbf{U}^T & \mathbf{0} 
   \end{bmatrix}  
   ,
\end{equation}
where $\mathbf{H}$ is Hermitian, meaning that it can be employed as a Hamiltonian in QC simulations. This step concludes the process of mapping the elastic wave equation with heterogeneous medium parameters to the Schr\"{o}dinger equation. It produces a Hamiltonian with two entries per row (2-sparse) for a finite-difference scheme of second-order accuracy in the 1-D case. We discuss higher space dimensions in section \ref{S:Discussion}.

\section{Implementation and experiments on quantum simulators and computers}\label{S:Implementation}

The transformations discussed in the previous sections are pre-processing steps executed on a classical computer.
In the following, we present the implementation of the transformed elastic wave equation on both an actual QC and an ideal (error-free) quantum simulator.
A quantum simulator is a software running on classical computers that emulates the behavior of actual QCs.
We create a quantum circuit that defines all operations that must be physically conducted on qubits during the circuit's runtime.
All of these operations, except for the measurements of the qubit states, are necessarily unitary.
Hence, the quantum circuit itself can be represented by a single unitary matrix.

The quantum circuit that solves the elastic wave equation through quantum time evolution is composed of 3 sub-circuits;
(1) one that prepares the initial condition $\ket{\phi(0)}$,
(2) one that performs the quantum simulation through implementing the Hamiltonian $\mathbf{H}$,
and (3) one that utilises the evolved state $\ket{\phi(t)}$.
This utilization can, for example, be a measurement of the qubits yielding binary results.
In this section, we employ the open-source software development kit \textsc{Qiskit} by IBM Research \citep{anis_qiskit_2021}
to create and run the quantum circuit on quantum simulators and actual QCs. The representation of the initial conditions $\ket{\phi(0)}$ on a QC requires normalisation,
\begin{equation}
    \ket{\psi(0)} = \frac{\ket{\phi(0)}}{||\ket{\phi(0)}||}\,,
\end{equation}
because quantum states must have a unit norm. This normalization has to be undone at the end of the simulation via
\begin{equation}
    \ket{\phi(t)} = \ket{\psi(t)} ||\ket{\phi(0)}||\,.
\end{equation}
Due to the linearity of (\ref{eq: quantum evolution}), no information is lost in this procedure.
We begin the simulation procedure with the initialization of an $n$-qubit quantum register,
\(\ket{00..00} = \ket{0}\otimes \dots \otimes \ket{0} \in \mathbb{C}^{2^{n}}\),
using the \textsc{Qiskit} constructor \texttt{qc = QuantumCircuit(n)}.
We must then apply the initialisation unitary transformation $\mathbf{V}$ to obtain the initial state,
\begin{equation}\label{eq: prepare initial state}
    \ket{\psi(0)} = \mathbf{V}\ket{00..00}\,.
\end{equation}
This is done by the \textsc{Qiskit} method \texttt{qc.prepare\_state(psi0)}.
Constructing and applying the quantum circuit corresponding to $\mathbf{V}$ can be challenging
\citep[e.g.,][]{kak1999initialization}.
However, in the context of seismic applications and using the above-derived transformation $\mathbf{T}$,
the initial state $\ket{\psi(0)}$ is sparse and can be initialised efficiently
as discussed in appendix \ref{A:Sparse Hamiltonian simulation}. Next, the quantum wave simulation circuit is realized by appending the time-evolution operator
\begin{equation}\label{eq: time-evolution operator}
    \mathbf{U}(t) = e^{-i\mathbf{H}t},
\end{equation}
to the quantum circuit, yielding
\begin{equation}\label{eq: time evolution}
     \ket{\psi(t)}  = \mathbf{U}(t) \ket{\psi(0)}.
\end{equation}
We currently perform the time evolution, i.e., the action of $\mathbf{U}$,
using the \textsc{Qiskit} class \texttt{PauliEvolutionGate}.
It implements the scaling and squaring method \citep{almohy2010scaling}.
The unitary time evolution for a particular evolution time, $\mathbf{U}(t)$, is constructed classically and decomposed into quantum gates. These gates are implemented as part of the overall quantum circuit. Therefore, the time evolution is equivalent to a single matrix-vector product between the time evolution operator $\mathbf{U}(t)$ and the initial state $\ket{\psi(0)}$.
We have chosen this method because it is implementable on today's freely accessible QCs, such as IBM Brisbane,
with a moderate circuit overhead for small-scale problems.
For large-scale problems, which will only become tractable on QCs with higher qubit counts and better error correction,
this approach becomes inapplicable due to the exponential scaling;
for $n$ qubits, $\mathbf{U}$ is a $2^n\times 2^n$ dense matrix that needs to be stored and applied to $\ket{\psi(0)}$.
As already explained in section \ref{S:Hamiltonian simulation},
the action of $\mathbf{U}$ must be carried out only approximately to obtain a scalable approach.
Appendix \ref{A:Sparse Hamiltonian simulation} discusses scaling if the best available approaches are applied to our settings.

Finally, the resulting quantum state $\ket{\psi(t)}$ is utilized further within the QC or is read out using quantum state tomography (QST) \citep{smolin2012efficient}.
Here, we opt for the latter to compare the QC simulation with the solution obtained on a classical computer. As discussed shortly, this approach is not practical for larger problems but is used here to validate the theory and implementation.

The squared entries $\{|\psi_i(t)|^2\}_{i=1}^{2^n}$ define a probability mass function which can be sampled by running and measuring the circuit multiple times. QST allows us to go beyond sampling the probability mass function but instead infer the actual amplitudes $\{\psi_i(t)\}_{i=1}^{2^n}$. It includes two steps: Firstly, estimating the expectation value of $3^n$ Pauli observables, which in our case reduces to $2^n$ because $\ket{\psi(t)}$ is purely real. Secondly, after all measurements are acquired, they are used in a least-squares inversion process to fit the amplitudes of $\ket{\psi(t)}$ by determining the density matrix $\ket{\psi(t)}\bra{\psi(t)}$ and its eigenvectors and eigenvalues \citep{smolin2012efficient}. Finally, it should be noted that the error in the estimated amplitudes $\{\psi^{est.}_i(t)\}_{i=1}^{2^n}$, given by $\epsilon = |\psi_i(t)-\psi^{est.}_i(t)|$, decreases up to prefactors with $1/\sqrt{S}$, where $S$ is the number of samples.

QST is prohibitively expensive for large problems, due to the exponential, here $3^n$ $(2^n)$ number of observables that need to be measured for a general complex (real) state. Additionally, to obtain a single sample with respect to an observable, a quantum measurement needs to be performed, which collapses the state \citep{nielsen2010quantum}, therefore requiring a new HS run for each sample.
Reading out the final state therefore requires exponentially many HS passes, nullifying any speedup from HS. We discuss this problem and perspectives to resolve it in the context of imaging in section \ref{S:Discussion}.

In summary, the quantum time evolution circuit proceeds as follows:
\begin{enumerate}
    \item Initialization of the quantum circuit with the appropriate number of bits, each in $\ket{0}$ state.
    \item Preparation of the state corresponding to the initial wave field per \eqref{eq: prepare initial state}.
    \item Time-evolution \eqref{eq: time evolution} via application of the time-evolution operator \eqref{eq: time-evolution operator}.
    \item Quantum state tomography (QST) via repeated HS runs and measurements.
\end{enumerate}
Details on the implemented quantum circuit using \textsc{Qiskit} can be found in Appendix \ref{A:Quantum Circuit Implementation}.

The quantum state tomography output, \(\ket{\psi(t)}\), represents the quantum state after the simulation.
To retrieve the classical wave field $[\mathbf{u}, \mathbf{v}]$, we reverse the normalization, apply the inverse transform $\mathbf{T}^{-1}$, and subsequently apply $\mathbf{M}^{-1/2}$.

\begin{figure}
    \includegraphics[width=1\linewidth]{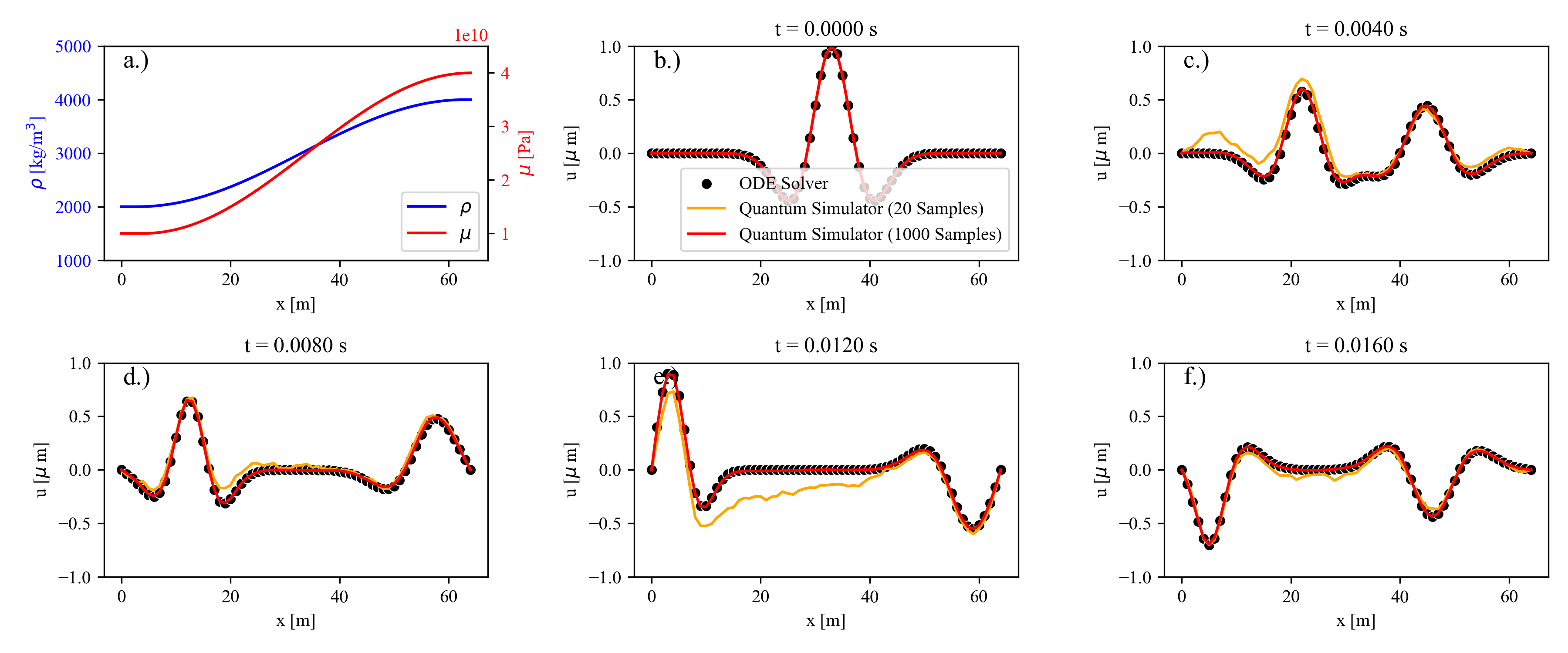}
    \caption{Comparison of elastic wave field simulations using a classical ODE solver (black) and two ideal quantum simulators with 20 samples (orange) and 1000 samples (red) used in the respective QST for a 128 grid-point problem. (a) Heterogeneous distributions of density and elastic modulus. (b-e) wave field at selected times.}
    \label{fig: fordward_sim_a}
\end{figure}

We use a QC simulator which is run on a classical computer to investigate the correctness of our algorithm. There are two types of simulators: one where artificial noise is added to simulate the real behavior of QCs and one which is noise-free. We opt for the latter to test the theory. Therefore, any divergence from the ground truth is attributed to finite sampling noise in the QST procedure. Only in the limit of infinitely many samples, QST reproduces the actual quantum state. 
As the numerical ground truth, we use an ordinary differential equation (ODE) solver based on an explicit Runge-Kutta method \citep[e.g.,][]{dormand1980family, shampine1986some} to directly integrate equation \eqref{eq: first order}.
We use Dirichlet boundary conditions, a uniform space discretization $dx$ of 1 m, and monotonically increasing density and elastic modulus distributions.
Since the quantum formalism is isomorphic to the discretized classical one, the HS should reproduce this solution, including finite-difference errors such as e.g. numerical dispersion. 
Figure \ref{fig: fordward_sim_a} displays the simulation results of the classical ground truth (black circles) and of quantum simulators using 20 (orange) and 1000 (red) samples for the QST, respectively.
We use a relative $L_2$ error (RL2) of the wave field amplitudes $u$ as a comparison metric calculated as
\begin{equation}
    E(t) = \frac{\sqrt{\sum_{i}^{N} (u_i(t) - u^{g}_i(t))^2}}{\sqrt{\sum_{i}^{N} u^{g}_i(t)^2}},
\end{equation}
with $u^g$ the ground truth solution obtained with the ODE solver.
For the time steps shown in Figure \ref{fig: fordward_sim_a}, the RL2 error is roughly 20 \% and 4 \% when $20$ and $1000$ samples are used, respectively.
Including even more samples would reduce the error further with a scaling of $1/\sqrt{S}$, as discussed above.
Because we use a matrix exponential scheme that evolves the state with a single matrix application (see Section \ref{S:Implementation}), the error does not accumulate over time but exclusively results from the QST and is determined by the statistics of the probability mass function of the state at the particular time point. However, for more elaborate time evolution algorithms that can be implemented on future QCs, such as the one discussed in Appendix \ref{A:Sparse Hamiltonian simulation}, the error will increase over time.

\begin{figure}
    \includegraphics[width=1\linewidth]{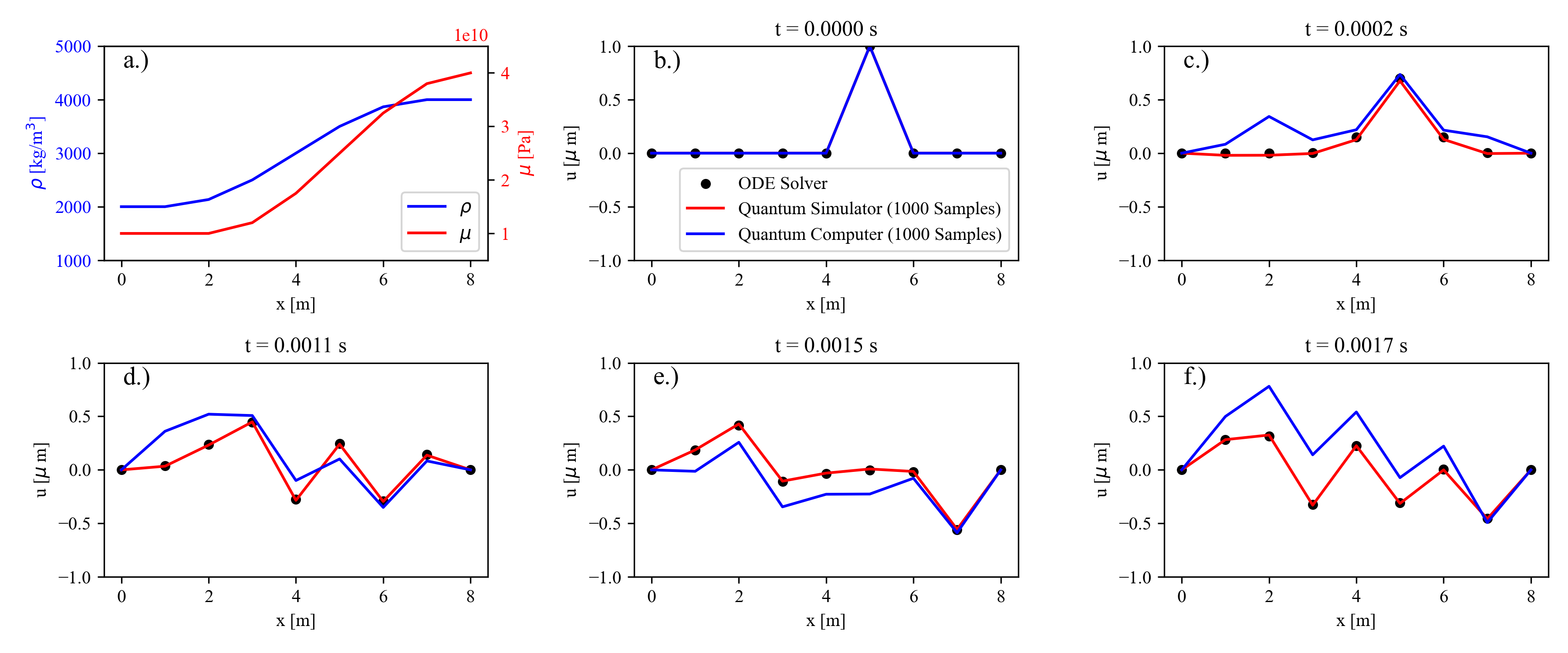}
    \caption{Comparison of elastic wave field simulations using a classical ODE solver (black), an ideal quantum simulator (red), and the quantum computer IBM Brisbane (blue) with 1000 samples used in the QST for a 7 grid-point problem. (a) Heterogeneous distributions of density and elastic modulus. (b-e) wave field at selected times.}
    \label{fig: fordward_sim_b}
\end{figure}

Finally, we solve the problem on the real QC IBM Brisbane, requiring us to scale the problem down to $7$ grid points, as discussed shortly. The results are displayed in Figure \ref{fig: fordward_sim_b}. We observe qualitative agreement between the numerical ground truth and the simulation on IBM Brisbane, however with RL2 errors of up to 60 \%.
These errors can be attributed to the high decoherence and noise rates on real QCs, as quantum states are extremely fragile and hard to isolate from the environment. We measure each quantum circuit with 1000 samples, reducing the RL2 contribution of the QST to approximately 3 \%, which is far below the hardware noise level that dominates the error. Again, the error does not accumulate over time, as discussed above, but rather depends on the quantum circuit for a particular time step and its hardware implementation.

When a quantum algorithm is executed, a sequence of quantum gates is applied to each qubit as discussed in section \ref{S:Quantum Computing}.
The number of gates applied to the qubits is known as the circuit depth.
Each gate application has a certain success rate, the probability that the gate operation is executed correctly.
IBM Brisbane's gates have an error per layered gate for a 100-qubit chain (EPLG) of 1.9 \%, equaling an approximate average 5-qubit layer success rate of 99.9 \% (\cite{gambetta2020ibm, collins2021ibm}).
For the 7 grid point problem, our implementation yields an average circuit depth of 870 quantum gates with 5 qubits,
which results roughly in a probability of $0.999^{870} \approx 0.42$
for a successful execution of the whole circuit.
This indeed reflects approximately the error level of 60 \% in our simulation.
Increasing the number of grid points will increase the circuit depth and hence the error beyond the noise level, currently limiting the problem size to 7 grid points. This is not a limit imposed by the presented algorithm, but a limitation of currently available hardware.

The error characteristics of QCs vary substantially per hardware and problem.
Different quantum processing units have different interaction topographies, implemented gate sets, and gate error rates of individual qubits and require continuous, often daily, re-calibration which changes the error characteristics. Because the problem specifics such as the number of grid points, evolution time, material parameters, initial conditions, or boundary conditions alter the quantum circuit, they influence the error. However, since currently the error mostly depends on the hardware which is calibrated and updated rapidly, a detailed problem-specific error analysis is not meaningful at this stage.

While currently HS on QCs is still in the proof-of-principle state, improving the reliability of quantum chips, in combination with more advanced error mitigation methods, is a vivid research field. It is expected that substantial improvements will be achieved in the coming years and decades (\cite{gambetta2020ibm, kim2023evidence}). Additionally, it has recently been demonstrated that HS is tractable on noisy QCs (\cite{kikuchi2023realization}), which may open avenues for applications of HS in the short term.

\section{Discussion}\label{S:Discussion}

In the following paragraphs, we provide additional details concerning potential computational speedups, the utilization of the final state, a continuous version of the above-described transformations in the context of alternative numerical methods, and future extensions of the 1-D elastic concept.

\subsection{Exponential quantum speedup}

Identifying the boundary beyond which QCs provide a computational advantage over their classical counterparts for a particular problem is a non-trivial task and subject to intensive theoretical \citep[e.g.,][]{grover1996fast,shor_polynomial-time_1997,liu_rigorous_2021,gyurik_towards_2022,franca_game_2022,hibat-allah_framework_2023,babbush2023exponential} as well as experimental research \citep[e.g.,][]{maslov_quantum_2021,liu_closing_2021,sanders_quantum_2021,bulmer_boundary_2022,huang_quantum_2022,madsen_quantum_2022}. Establishing practical results based on theory largely depends on the anticipated emergence of error-corrected quantum chips, as opposed to today's NISQ machines \citep{preskill2018quantum}.

The first promise of quantum computing concerns the storage size. For classical computational methods, the number of bits required to store the entire wave field scales linearly with the number of grid points. In contrast, for quantum computational methods like the one discussed here, the number of required qubits scales only logarithmically because, reciprocally, the quantum state space grows exponentially with the number of qubits. For instance, assume a discretized wave field with two real-valued parameters per grid point, such as displacement and velocity, and Earth volume $1.083 \times 10^{12}\ \textrm{km}^3$. Representing a full wave field in $\textrm{mm}^3$ resolution requires merely 101 qubits.

The second promise is runtime speedup. As discussed in Section \ref{S:Implementation}, the time evolution method implemented in our QC experiment has been chosen because it has a small circuit overhead, allowing for implementation on current hardware. This formulation does not provide any speedup over classical computers as it scales exponentially in the number of qubits. However, as detailed in Appendix \ref{A:Sparse Hamiltonian simulation}, the best quantum algorithms for HS have a runtime complexity that is polynomial in $n$ and poly-logarithmic in $N$ (\cite{low2019hamiltonian, babbush2023exponential}). We analyze those algorithms only theoretically as they are not implementable on current hardware due to the large circuit overhead.

In contrast to the best HS algorithms, any classical algorithm necessarily has at least exponential complexity in $n$ and polynomial in $N$ \citep{babbush2023exponential}. In the foreseeable future, the specific $N$ for which the quantum approach becomes faster might be above the limit given by the available hardware. However, the promise is highly enticing in the long term. 

\subsection{The read-out problem}

As discussed in section \ref{S:Implementation}, reading out the full state with QST scales polynomially in $N$ and exponentially in $n$. Hence, the read-out would nullify the exponential speedup. \cite{babbush2023exponential} show that it is possible to read out certain properties, such as kinetic energy or potential energy of a subdomain of the model, in polynomial time, suggesting that the exponential speedup can be preserved in such cases. A similar logic will be required to preserve the exponential speedup for seismological purposes,
e.g., by computing a misfit function on the fly instead of reading out the complete time evolution of the wave field. This topic will be the subject of a follow-up publication.

\subsection{Continuous transformation and the use of other numerical schemes}\label{SS:continuous}

In section \ref{S:Theory}, we performed the transformation of the elastic wave equation to the Schrödinger equation in the discrete domain in the interest of conciseness. A conceptually similar transformation can be designed in the continuous domain, thereby enabling the use of alternative numerical methods, including finite-element or spectral-element schemes. Furthermore, the continuous approach illuminates how transformations must be designed to achieve self-adjointness with respect to the natural inner product in quantum physics, loosely referred to as the quantum inner product in the following. To understand this, we consider the wave operator for the 1-D elastic wave equation
\begin{equation}
    \mathcal{L} := \partial_x[\mu\partial_x]\,.
\end{equation}
We omit the dependence on space, $x$ for brevity. Following the approach in section \ref{S:Theory}, we define a transformation utilising the square root of density $\rho^{1/2}$, resulting in the density-transformed wave field $\tilde{u} = \rho^{1/2} u$ and a density-transformed wave operator
\begin{equation}\label{eq: density transformed waveop}
     \mathcal{L}_{\rho}:=\rho^{-1/2}\partial_x[\mu \partial_x]\rho^{-1/2}\,.
\end{equation}
We then transform the continuous wave equation to a first-order system
\begin{equation} \label{eq: first order}
    \partial_t \begin{bmatrix}
      \tilde{u} \\ 
      \tilde{v}
   \end{bmatrix}= 
   \begin{bmatrix}
      0 & 1\\
      \mathcal{L}_{\rho} & 0
   \end{bmatrix}
   \begin{bmatrix}
      \tilde{u} \\ 
      \tilde{v}
   \end{bmatrix} 
   =: 
   \mathcal{Q} \begin{bmatrix}
      \tilde{u} \\ 
      \tilde{v}
   \end{bmatrix},
\end{equation}
where $\tilde{v} = \partial_t \tilde{u} = \rho^{1/2} \partial_t u \in \mathbb{R}$ are the density-transformed velocities.
We now aim to construct an invertible transformation $\mathcal{T}$ that maps equation \eqref{eq: first order} to a continuous Schrödinger equation.
Mathematically, this means that $H = i\mathcal{T}\mathcal{Q}\mathcal{T}^{-1}$ has to be self-adjoint under the quantum inner product
\begin{equation}\label{eq: inner product augm}
 \left\langle\begin{bmatrix}
      a \\ 
      b
   \end{bmatrix}\middle\vert
   \begin{bmatrix}
    c\\ 
      d
   \end{bmatrix}\right\rangle = \braket{a,c} + \braket{b,d} 
   =  \int_0^W dx \, a^*(x)c(x) + \int_0^W dx\, b^*(x)d(x).
\end{equation}
In appendix \ref{A:Continuous transformation}, we show that there exists such a transformation that is invertible. The resulting Hamiltonian is given by
\begin{equation}
    \mathcal{H}
= i\begin{bmatrix}
  0 & \mu^{1/2}\partial_x\rho^{-1/2}\\
  \rho^{1/2}\partial_x\mu^{-1/2} & 0
\end{bmatrix},
\end{equation}
and the  transformed wave field vector is identified as $\ket{\phi} = \mathcal{T}[\tilde{u},\tilde{v}]^T$.
This leads to a continuous Schrödinger equation
\begin{equation}\label{eq: quantum evolution 2}
    i \partial_t \ket{\phi(t)} = \mathcal{H} \ket{\phi(t)},
\end{equation}
subject to the transformed initial conditions
\begin{equation}\label{eq: transformed initial conds}
   \ket{ \phi(0)} = \mathcal{T} \begin{bmatrix}
      \rho^{1/2}u_0 \\ 
      \rho^{1/2}v_0
   \end{bmatrix}.
\end{equation}
It is now possible to discretize the continuous Schrödinger equation directly by discretizing $\mu^{1/2}\partial_x\rho^{-1/2}$ with any numerical method, provided that the boundary conditions can be satisfied.

\subsection{Extensions and limitations}\label{SS:extensions}

The work presented here is merely a first step toward QC wave simulations in geophysical applications. Generalizing to higher space dimensions will require a similar procedure as above to map the augmented wave operator $\mathcal{Q}$ or its discretized version $\mathbf{Q}$ to the Hamiltonian operator, ensuring self-adjointness under the quantum inner product. A bigger challenge may be the incorporation of visco-elastic attenuation because HS is inherently unitary and hence can only simulate systems that conserve energy. Another open problem is the incorporation of sources, which will be the topic of a follow-up study.

\subsection{Relation to other quantum algorithms with potential for seismological research}\label{ss: relation to other algs}

In this article, the QC was used as a semi-analog simulator through HS. There exist other families of algorithms that have the potential to be harnessed in seismology. Among the most prominent are variational quantum algorithms (\cite{cerezo2021variational}) which use a hybrid quantum-classical optimization procedure to find eigenvalues and eigenvectors (\cite{abrams1999quantum, kirby2021variational}) or solve linear systems (\cite{pellow2021comparison}).  The former may have applications in solving the wave equation in the frequency domain. The latter has been used to solve linear partial differential equations, including the wave equation (\cite{trahan2023variational}), by successively solving the linear system of equations that arises in implicit time stepping schemes. While their computational benefits in comparison to classical solutions and HS still need to be worked out, they bring the advantage of being executable on today's and near-term quantum chips of the NISQ era (\cite{cerezo2021variational}).

Again looking ahead into the potential future of error-corrected quantum chips, \cite{harrow2009quantum}  have proposed a quantum algorithm, known as the HHL algorithm, that is capable of solving sparse linear systems with an exponential speedup over classical computers. Such algorithms can eventually be harnessed in a discrete-time integration scheme similar to \cite{trahan2023variational} or in linearized inverse problems (\cite{wiebe2012quantum, schuld2016prediction}). Furthermore, despite concerns over vanishing gradients issues (\cite{mcclean2018barren,wang2021noise}), quantum machine learning (\cite{biamonte2017quantum, melnikov2023quantum}) is an emerging field with potential application in seismological research.

Finally, there are quantum annealers (QAs) or adiabatic quantum computers (\cite{albash2018adiabatic}), which differ from gate-based QCs. QAs are capable of solving quadratic optimization problems and their application in geoscience has been investigated in a handful of studies (\cite{o2018approach, golden2021pre, souza2022application, dukalski2023quantum}). It should be noted that there is controversy about whether or not QA can achieve the same speedup as gate-based quantum computation in real-world contexts (\cite{van2001powerful, aharonov2008adiabatic, albash2018demonstration, villanueva2023adiabatic}).

\section{Conclusions and outlook}\label{S:Conclusions}

We presented a quantum computing concept for 1-D elastic wave propagation through heterogeneous media. Our approach involves (1) a finite-difference approximation of the wave equation, (2) a sparsity-preserving transformation to a Schr\"{o}dinger equation, and (3) a direct simulation of the Schr\"{o}dinger equation on a gate-based quantum computer. 

The implementation of our approach on an error-free quantum simulator produces nearly exact results and serves as the foundation of wave field simulations with small numbers of grid points on the real quantum computer IBM Brisbane. While the simulation results qualitatively agree with the error-free version, they are contaminated by errors related to quantum decoherence and noise, which are typical issues for today's quantum computers.

A continuous version of the discrete transformation to the Schr\"{o}dinger equation enables the modification of our method based on finite differences, including the use of other spatial discretization schemes, such as the spectral-element method. Finally, assuming that error-corrected quantum chips will become available, we discuss that the best quantum algorithms can solve the elastic wave equation with exponential speedup over classical solvers.

With the FWI as an end goal, our ongoing research focuses on incorporating sources, measuring misfit functions, extending the formalism to higher space dimensions, including visco-elastic attenuation, and computing gradients.

\begin{acknowledgments}
We would like to thank the other members of the ETH Seismology \& Wave Physics Group for fruitful discussions and inspiration. We would also like to acknowledge Shahpoor Moradi, Alexandre Souza, and one anonymous reviewer for insightful feedback. All codes used in this work are available at \url{https://github.com/malteschade/Quantum-Wave-Equation-Solver}.
\end{acknowledgments}

\noindent \textbf{Data statement}: No data have been used for this research.

\noindent \textbf{Author contribution statement}: theory and study conception: C. Bösch; implementation and execution on the quantum computer: M. Schade; quantum computing introduction: V. Hapla; analysis and interpretation of results: M. Schade, C. Bösch, V. Hapla, A. Fichtner; manuscript preparation: M. Schade, C. Bösch, V. Hapla, A. Fichtner. All authors reviewed the results and approved the final version of the manuscript.


\bibliography{biblio}

\begin{thebibliography}{}

\bibitem[\protect\citeauthoryear{Abrams and Lloyd}{Abrams and Lloyd}{1999}]{abrams1999quantum}
Abrams, D.~S. and S.~Lloyd (1999).
\newblock Quantum algorithm providing exponential speed increase for finding eigenvalues and eigenvectors.
\newblock {\em Physical Review Letters\/}~{\em 83\/}(24), 5162.

\bibitem[\protect\citeauthoryear{Adedoyin, Ambrosiano, Anisimov, Casper, Chennupati, Coffrin, Djidjev, Gunter, Karra, and Lemons}{Adedoyin et~al.}{2022}]{quantum_algorithms_beginners_2022}
Adedoyin, A., J.~Ambrosiano, P.~Anisimov, W.~Casper, G.~Chennupati, C.~Coffrin, H.~Djidjev, D.~Gunter, S.~Karra, and N.~Lemons (2022).
\newblock Quantum algorithm implementations for beginners.
\newblock {\em ACM Transactions on Quantum Computing\/}~{\em 3\/}(4), 18:1--18:92.

\bibitem[\protect\citeauthoryear{Aharonov, Van~Dam, Kempe, Landau, Lloyd, and Regev}{Aharonov et~al.}{2008}]{aharonov2008adiabatic}
Aharonov, D., W.~Van~Dam, J.~Kempe, Z.~Landau, S.~Lloyd, and O.~Regev (2008).
\newblock Adiabatic quantum computation is equivalent to standard quantum computation.
\newblock {\em SIAM review\/}~{\em 50\/}(4), 755--787.

\bibitem[\protect\citeauthoryear{Al-Mohy and Higham}{Al-Mohy and Higham}{2010}]{almohy2010scaling}
Al-Mohy, A.~H. and N.~J. Higham (2010).
\newblock A new scaling and squaring algorithm for the matrix exponential.
\newblock {\em SIAM Journal on Matrix Analysis and Applications\/}~{\em 31\/}(3), 970--989.

\bibitem[\protect\citeauthoryear{Albash and Lidar}{Albash and Lidar}{2018a}]{albash2018adiabatic}
Albash, T. and D.~A. Lidar (2018a).
\newblock Adiabatic quantum computation.
\newblock {\em Reviews of Modern Physics\/}~{\em 90\/}(1), 015002.

\bibitem[\protect\citeauthoryear{Albash and Lidar}{Albash and Lidar}{2018b}]{albash2018demonstration}
Albash, T. and D.~A. Lidar (2018b).
\newblock Demonstration of a scaling advantage for a quantum annealer over simulated annealing.
\newblock {\em Physical Review X\/}~{\em 8\/}(3), 031016.

\bibitem[\protect\citeauthoryear{Aleksandrowicz, Alexander, Barkoutsos, Bello, Ben-Haim, Bucher, Cabrera-Hern{\'a}ndez, Carballo-Franquis, Chen, Chen, et~al.}{Aleksandrowicz et~al.}{2019}]{anis_qiskit_2021}
Aleksandrowicz, G., T.~Alexander, P.~Barkoutsos, L.~Bello, Y.~Ben-Haim, D.~Bucher, F.~J. Cabrera-Hern{\'a}ndez, J.~Carballo-Franquis, A.~Chen, C.-F. Chen, et~al. (2019).
\newblock Qiskit: An open-source framework for quantum computing.

\bibitem[\protect\citeauthoryear{Babbush, Berry, Kothari, Somma, and Wiebe}{Babbush et~al.}{2023}]{babbush2023exponential}
Babbush, R., D.~W. Berry, R.~Kothari, R.~D. Somma, and N.~Wiebe (2023).
\newblock Exponential quantum speedup in simulating coupled classical oscillators.
\newblock {\em arXiv preprint arXiv:2303.13012\/}.

\bibitem[\protect\citeauthoryear{Benenti, Casati, Rossini, and Strini}{Benenti et~al.}{2019}]{benenti2019principles}
Benenti, G., G.~Casati, D.~Rossini, and G.~Strini (2019).
\newblock {\em Principles of quantum computation and information: a comprehensive textbook}.
\newblock World Scientific.

\bibitem[\protect\citeauthoryear{Benenti, Casati, and Strini}{Benenti et~al.}{2004}]{benenti2004principles}
Benenti, G., G.~Casati, and G.~Strini (2004).
\newblock {\em Principles of quantum computation and information-volume I: Basic concepts}.
\newblock World scientific.

\bibitem[\protect\citeauthoryear{Berry, Ahokas, Cleve, and Sanders}{Berry et~al.}{2007}]{berry2007efficient}
Berry, D.~W., G.~Ahokas, R.~Cleve, and B.~C. Sanders (2007).
\newblock Efficient quantum algorithms for simulating sparse hamiltonians.
\newblock {\em Communications in Mathematical Physics\/}~{\em 270}, 359--371.

\bibitem[\protect\citeauthoryear{Berry, Childs, Cleve, Kothari, and Somma}{Berry et~al.}{2015}]{berry2015truncated}
Berry, D.~W., A.~M. Childs, R.~Cleve, R.~Kothari, and R.~D. Somma (2015).
\newblock Simulating {Hamiltonian} dynamics with a truncated {Taylor} series.
\newblock {\em Phys. Rev. Lett.\/}~{\em 114}, 090502--1--090502--5.

\bibitem[\protect\citeauthoryear{Biamonte, Wittek, Pancotti, Rebentrost, Wiebe, and Lloyd}{Biamonte et~al.}{2017}]{biamonte2017quantum}
Biamonte, J., P.~Wittek, N.~Pancotti, P.~Rebentrost, N.~Wiebe, and S.~Lloyd (2017).
\newblock Quantum machine learning.
\newblock {\em Nature\/}~{\em 549\/}(7671), 195--202.

\bibitem[\protect\citeauthoryear{Bozda\u{g}, Peter, Lefebvre, Komatitsch, Tromp, Hill, Podhorszki, and Pugmire}{Bozda\u{g} et~al.}{2016}]{Bozdag_2016}
Bozda\u{g}, E., D.~Peter, M.~Lefebvre, D.~Komatitsch, J.~Tromp, J.~Hill, N.~Podhorszki, and D.~Pugmire (2016).
\newblock {Global adjoint tomography: First-generation model}.
\newblock {\em Geophys. J. Int.\/}~{\em 207}, 1739--1766.

\bibitem[\protect\citeauthoryear{Bravyi, Cross, Gambetta, Maslov, Rall, and Yoder}{Bravyi et~al.}{2023}]{bravyi2023high}
Bravyi, S., A.~W. Cross, J.~M. Gambetta, D.~Maslov, P.~Rall, and T.~J. Yoder (2023).
\newblock High-threshold and low-overhead fault-tolerant quantum memory.
\newblock {\em arXiv preprint arXiv:2308.07915\/}.

\bibitem[\protect\citeauthoryear{Bulmer, Bell, Chadwick, Jones, Moise, Rigazzi, Thorbecke, Haus, Van~Vaerenbergh, and Patel}{Bulmer et~al.}{2022}]{bulmer_boundary_2022}
Bulmer, J.~F., B.~A. Bell, R.~S. Chadwick, A.~E. Jones, D.~Moise, A.~Rigazzi, J.~Thorbecke, U.-U. Haus, T.~Van~Vaerenbergh, and R.~B. Patel (2022).
\newblock The boundary for quantum advantage in gaussian boson sampling.
\newblock {\em Science advances\/}~{\em 8\/}(4), eabl9236.

\bibitem[\protect\citeauthoryear{Cerezo, Arrasmith, Babbush, Benjamin, Endo, Fujii, McClean, Mitarai, Yuan, Cincio, et~al.}{Cerezo et~al.}{2021}]{cerezo2021variational}
Cerezo, M., A.~Arrasmith, R.~Babbush, S.~C. Benjamin, S.~Endo, K.~Fujii, J.~R. McClean, K.~Mitarai, X.~Yuan, L.~Cincio, et~al. (2021).
\newblock Variational quantum algorithms.
\newblock {\em Nature Reviews Physics\/}~{\em 3\/}(9), 625--644.

\bibitem[\protect\citeauthoryear{Childs, Maslov, Nam, Ross, and Su}{Childs et~al.}{2018}]{childs2018toward}
Childs, A.~M., D.~Maslov, Y.~Nam, N.~J. Ross, and Y.~Su (2018).
\newblock Toward the first quantum simulation with quantum speedup.
\newblock {\em Proceedings of the National Academy of Sciences\/}~{\em 115\/}(38), 9456--9461.

\bibitem[\protect\citeauthoryear{Collins and Easterly}{Collins and Easterly}{2021}]{collins2021ibm}
Collins, H. and K.~Easterly (2021).
\newblock Ibm unveils breakthrough 127-qubit quantum processor.
\newblock {\em IBM Newsroom\/}.

\bibitem[\protect\citeauthoryear{Costa, Jordan, and Ostrander}{Costa et~al.}{2019}]{costa2019quantum}
Costa, P.~C., S.~Jordan, and A.~Ostrander (2019).
\newblock Quantum algorithm for simulating the wave equation.
\newblock {\em Physical Review A\/}~{\em 99\/}(1), 012323.

\bibitem[\protect\citeauthoryear{Daley, Bloch, Kokail, Flannigan, Pearson, Troyer, and Zoller}{Daley et~al.}{2022}]{daley2022practical}
Daley, A.~J., I.~Bloch, C.~Kokail, S.~Flannigan, N.~Pearson, M.~Troyer, and P.~Zoller (2022).
\newblock Practical quantum advantage in quantum simulation.
\newblock {\em Nature\/}~{\em 607\/}(7920), 667--676.

\bibitem[\protect\citeauthoryear{Dhand and Sanders}{Dhand and Sanders}{2014}]{dhand_stability_2014}
Dhand, I. and B.~C. Sanders (2014).
\newblock Stability of the {Trotter}–{Suzuki} decomposition.
\newblock {\em Journal of Physics A: Mathematical and Theoretical\/}~{\em 47\/}(26), 265206.

\bibitem[\protect\citeauthoryear{Dormand and Prince}{Dormand and Prince}{1980}]{dormand1980family}
Dormand, J.~R. and P.~J. Prince (1980).
\newblock A family of embedded runge-kutta formulae.
\newblock {\em Journal of computational and applied mathematics\/}~{\em 6\/}(1), 19--26.

\bibitem[\protect\citeauthoryear{Dukalski, Rovetta, van~der Linde, M{\"o}ller, Neumann, and Phillipson}{Dukalski et~al.}{2023}]{dukalski2023quantum}
Dukalski, M., D.~Rovetta, S.~van~der Linde, M.~M{\"o}ller, N.~Neumann, and F.~Phillipson (2023).
\newblock Quantum computer-assisted global optimization in geophysics illustrated with stack-power maximization for refraction residual statics estimation.
\newblock {\em Geophysics\/}~{\em 88\/}(2), V75--V91.

\bibitem[\protect\citeauthoryear{Fichtner}{Fichtner}{2010}]{fichtner2010full}
Fichtner, A. (2010).
\newblock {\em Full seismic waveform modelling and inversion}.
\newblock Springer Science \& Business Media.

\bibitem[\protect\citeauthoryear{França and Garcia-Patron}{França and Garcia-Patron}{2022}]{franca_game_2022}
França, D.~S. and R.~Garcia-Patron (2022).
\newblock A game of quantum advantage: Linking verification and simulation.
\newblock {\em Quantum\/}~{\em 6}, 753.

\bibitem[\protect\citeauthoryear{Gambetta}{Gambetta}{2020}]{gambetta2020ibm}
Gambetta, J. (2020).
\newblock Ibm’s roadmap for scaling quantum technology.
\newblock {\em IBM Research Blog (September 2020)\/}.

\bibitem[\protect\citeauthoryear{Gebraad and Fichtner}{Gebraad and Fichtner}{2023}]{gebraad_seamless}
Gebraad, L. and A.~Fichtner (2023).
\newblock Seamless {GPU} acceleration for {C++}‐based physics with the {Metal Shading Language} on {Apple}’s {M} series unified chips.
\newblock {\em Seismological Research Letters\/}~{\em 94\/}(3), 1670--1675.

\bibitem[\protect\citeauthoryear{Georgescu, Ashhab, and Nori}{Georgescu et~al.}{2014}]{georgescu2014quantum}
Georgescu, I.~M., S.~Ashhab, and F.~Nori (2014).
\newblock Quantum simulation.
\newblock {\em Reviews of Modern Physics\/}~{\em 86\/}(1), 153.

\bibitem[\protect\citeauthoryear{Gibney}{Gibney}{2019}]{gibney2019hello}
Gibney, E. (2019).
\newblock Hello quantum world! google publishes landmark quantum supremacy claim.
\newblock {\em Nature\/}~{\em 574\/}(7779), 461--463.

\bibitem[\protect\citeauthoryear{Gleinig and Hoefler}{Gleinig and Hoefler}{2021}]{gleinig2021efficient}
Gleinig, N. and T.~Hoefler (2021).
\newblock An efficient algorithm for sparse quantum state preparation.
\newblock In {\em Proceedings: 58th Design Automation Conference}, pp.\  433--438. IEEG.

\bibitem[\protect\citeauthoryear{Golden and O’Malley}{Golden and O’Malley}{2021}]{golden2021pre}
Golden, J.~K. and D.~O’Malley (2021).
\newblock Pre-and post-processing in quantum-computational hydrologic inverse analysis.
\newblock {\em Quantum Information Processing\/}~{\em 20\/}(5), 176.

\bibitem[\protect\citeauthoryear{Grover}{Grover}{1996}]{grover1996fast}
Grover, L.~K. (1996).
\newblock A fast quantum mechanical algorithm for database search.
\newblock In {\em Proceedings of the twenty-eighth annual ACM symposium on Theory of computing}, pp.\  212--219.

\bibitem[\protect\citeauthoryear{Gyurik, Cade, and Dunjko}{Gyurik et~al.}{2022}]{gyurik_towards_2022}
Gyurik, C., C.~Cade, and V.~Dunjko (2022).
\newblock Towards quantum advantage via topological data analysis.
\newblock {\em Quantum\/}~{\em 6}, 855.

\bibitem[\protect\citeauthoryear{Harrow, Hassidim, and Lloyd}{Harrow et~al.}{2009}]{harrow2009quantum}
Harrow, A.~W., A.~Hassidim, and S.~Lloyd (2009).
\newblock Quantum algorithm for linear systems of equations.
\newblock {\em Physical review letters\/}~{\em 103\/}(15), 150502.

\bibitem[\protect\citeauthoryear{Hatano and Suzuki}{Hatano and Suzuki}{2005}]{hatano2005finding}
Hatano, N. and M.~Suzuki (2005).
\newblock {\em Finding exponential product formulas of higher orders}, pp.\  37–68.
\newblock Springer.

\bibitem[\protect\citeauthoryear{Hibat-Allah, Mauri, Carrasquilla, and Perdomo-Ortiz}{Hibat-Allah et~al.}{2023}]{hibat-allah_framework_2023}
Hibat-Allah, M., M.~Mauri, J.~Carrasquilla, and A.~Perdomo-Ortiz (2023).
\newblock A framework for demonstrating practical quantum advantage: Racing quantum against classical generative models.

\bibitem[\protect\citeauthoryear{Huang, Broughton, Cotler, Chen, Li, Mohseni, Neven, Babbush, Kueng, and Preskill}{Huang et~al.}{2022}]{huang_quantum_2022}
Huang, H.-Y., M.~Broughton, J.~Cotler, S.~Chen, J.~Li, M.~Mohseni, H.~Neven, R.~Babbush, R.~Kueng, and J.~Preskill (2022).
\newblock Quantum advantage in learning from experiments.
\newblock {\em Science\/}~{\em 376\/}(6598), 1182--1186.

\bibitem[\protect\citeauthoryear{Igel, Djikpesse, and Tarantola}{Igel et~al.}{1996}]{Igel_1996}
Igel, H., H.~Djikpesse, and A.~Tarantola (1996).
\newblock Waveform inversion of marine reflection seismograms for {P} impedance and {P}oisson's ratio.
\newblock {\em Geophys. J. Int.\/}~{\em 124}, 363--371.

\bibitem[\protect\citeauthoryear{Igel, Mora, and Riollet}{Igel et~al.}{1995}]{Igel_1995}
Igel, H., P.~Mora, and B.~Riollet (1995).
\newblock {Anisotropic wave propagation through FD grids}.
\newblock {\em Geophysics\/}~{\em 60}, 1203--1216.

\bibitem[\protect\citeauthoryear{Jin, Liu, and Ma}{Jin et~al.}{2023}]{jin2023quantum}
Jin, S., N.~Liu, and C.~Ma (2023).
\newblock Quantum simulation of maxwell's equations via schr{\"o}odingersation.
\newblock {\em arXiv preprint arXiv:2308.08408\/}.

\bibitem[\protect\citeauthoryear{Jin, Liu, and Yu}{Jin et~al.}{2022}]{jin2022quantum}
Jin, S., N.~Liu, and Y.~Yu (2022).
\newblock Quantum simulation of partial differential equations via schrodingerisation: Technical details.
\newblock {\em arXiv preprint arXiv:2212.14703\/}.

\bibitem[\protect\citeauthoryear{Kak}{Kak}{1999}]{kak1999initialization}
Kak, S. (1999).
\newblock The initialization problem in quantum computing.
\newblock {\em Foundations of Physics\/}~{\em 29}, 267--279.

\bibitem[\protect\citeauthoryear{Kane and Lubensky}{Kane and Lubensky}{2014}]{kane2014topological}
Kane, C.~L. and T.~C. Lubensky (2014).
\newblock Topological boundary modes in isostatic lattices.
\newblock {\em Nature Physics\/}~{\em 10\/}(1), 39--45.

\bibitem[\protect\citeauthoryear{Kikuchi, Mc~Keever, Coopmans, Lubasch, and Benedetti}{Kikuchi et~al.}{2023}]{kikuchi2023realization}
Kikuchi, Y., C.~Mc~Keever, L.~Coopmans, M.~Lubasch, and M.~Benedetti (2023).
\newblock Realization of quantum signal processing on a noisy quantum computer.
\newblock {\em npj Quantum Information\/}~{\em 9\/}(1), 93.

\bibitem[\protect\citeauthoryear{Kim, Eddins, Anand, Wei, Van Den~Berg, Rosenblatt, Nayfeh, Wu, Zaletel, Temme, et~al.}{Kim et~al.}{2023}]{kim2023evidence}
Kim, Y., A.~Eddins, S.~Anand, K.~X. Wei, E.~Van Den~Berg, S.~Rosenblatt, H.~Nayfeh, Y.~Wu, M.~Zaletel, K.~Temme, et~al. (2023).
\newblock Evidence for the utility of quantum computing before fault tolerance.
\newblock {\em Nature\/}~{\em 618\/}(7965), 500--505.

\bibitem[\protect\citeauthoryear{Kirby and Love}{Kirby and Love}{2021}]{kirby2021variational}
Kirby, W.~M. and P.~J. Love (2021).
\newblock Variational quantum eigensolvers for sparse hamiltonians.
\newblock {\em Physical review letters\/}~{\em 127\/}(11), 110503.

\bibitem[\protect\citeauthoryear{Knill}{Knill}{2005}]{knill2005quantum}
Knill, E. (2005).
\newblock Quantum computing with realistically noisy devices.
\newblock {\em Nature\/}~{\em 434\/}(7029), 39--44.

\bibitem[\protect\citeauthoryear{Liu and Gu}{Liu and Gu}{2012}]{Liu_2012}
Liu, Q. and Y.~Gu (2012).
\newblock {Seismic imaging: from classical to adjoint tomography}.
\newblock {\em Tectonophysics\/}~{\em 566-567}, 31--66.

\bibitem[\protect\citeauthoryear{Liu, Arunachalam, and Temme}{Liu et~al.}{2021}]{liu_rigorous_2021}
Liu, Y., S.~Arunachalam, and K.~Temme (2021).
\newblock A rigorous and robust quantum speed-up in supervised machine learning.
\newblock {\em Nature Physics\/}~{\em 17\/}(9), 1013--1017.

\bibitem[\protect\citeauthoryear{Liu, Liu, Li, Fu, Yang, Song, Zhao, Wang, Peng, and Chen}{Liu et~al.}{2021}]{liu_closing_2021}
Liu, Y., X.~Liu, F.~Li, H.~Fu, Y.~Yang, J.~Song, P.~Zhao, Z.~Wang, D.~Peng, and H.~Chen (2021).
\newblock Closing the" quantum supremacy" gap: Achieving real-time simulation of a random quantum circuit using a new sunway supercomputer.
\newblock In {\em Proceedings of the {International} {Conference} for {High} {Performance} {Computing}, {Networking}, {Storage} and {Analysis}}, {SC} '21, New York, NY, USA, pp.\  1--12. Association for Computing Machinery.

\bibitem[\protect\citeauthoryear{Low and Chuang}{Low and Chuang}{2019}]{low2019hamiltonian}
Low, G.~H. and I.~L. Chuang (2019).
\newblock Hamiltonian simulation by qubitization.
\newblock {\em Quantum\/}~{\em 3}, 163.

\bibitem[\protect\citeauthoryear{Madsen, Laudenbach, Askarani, Rortais, Vincent, Bulmer, Miatto, Neuhaus, Helt, and Collins}{Madsen et~al.}{2022}]{madsen_quantum_2022}
Madsen, L.~S., F.~Laudenbach, M.~F. Askarani, F.~Rortais, T.~Vincent, J.~F. Bulmer, F.~M. Miatto, L.~Neuhaus, L.~G. Helt, and M.~J. Collins (2022).
\newblock Quantum computational advantage with a programmable photonic processor.
\newblock {\em Nature\/}~{\em 606\/}(7912), 75--81.

\bibitem[\protect\citeauthoryear{Maslov, Kim, Bravyi, Yoder, and Sheldon}{Maslov et~al.}{2021}]{maslov_quantum_2021}
Maslov, D., J.-S. Kim, S.~Bravyi, T.~J. Yoder, and S.~Sheldon (2021).
\newblock Quantum advantage for computations with limited space.
\newblock {\em Nature Physics\/}~{\em 17\/}(8), 894--897.

\bibitem[\protect\citeauthoryear{McClean, Boixo, Smelyanskiy, Babbush, and Neven}{McClean et~al.}{2018}]{mcclean2018barren}
McClean, J.~R., S.~Boixo, V.~N. Smelyanskiy, R.~Babbush, and H.~Neven (2018).
\newblock Barren plateaus in quantum neural network training landscapes.
\newblock {\em Nature communications\/}~{\em 9\/}(1), 4812.

\bibitem[\protect\citeauthoryear{Melnikov, Kordzanganeh, Alodjants, and Lee}{Melnikov et~al.}{2023}]{melnikov2023quantum}
Melnikov, A., M.~Kordzanganeh, A.~Alodjants, and R.-K. Lee (2023).
\newblock Quantum machine learning: from physics to software engineering.
\newblock {\em Advances in Physics: X\/}~{\em 8\/}(1), 2165452.

\bibitem[\protect\citeauthoryear{Moczo, Kristek, and G{\'a}lis}{Moczo et~al.}{2014}]{moczo2014finite}
Moczo, P., J.~Kristek, and M.~G{\'a}lis (2014).
\newblock {\em The finite-difference modelling of earthquake motions: Waves and ruptures}.
\newblock CUP.

\bibitem[\protect\citeauthoryear{Montanaro}{Montanaro}{2016}]{montanaro2016quantum}
Montanaro, A. (2016).
\newblock Quantum algorithms: An overview.
\newblock {\em npj Quantum Information\/}~{\em 2\/}(1), 1--8.

\bibitem[\protect\citeauthoryear{Moradi, Trad, and Innanen}{Moradi et~al.}{2018}]{moradi2018quantum}
Moradi, S., D.~Trad, and K.~A. Innanen (2018).
\newblock Quantum computing in geophysics: Algorithms, computational costs, and future applications.
\newblock In {\em SEG International Exposition and Annual Meeting}, pp.\  SEG--2018. SEG.

\bibitem[\protect\citeauthoryear{Moradi, Trad, and Innanen}{Moradi et~al.}{2019}]{moradi1019quantum}
Moradi, S., D.~Trad, and K.~A. Innanen (2019, 12).
\newblock When quantum computers arrive on seismology’s doorstep.
\newblock {\em Canadian Journal of Exploration Geophysics\/}~{\em 44}, 1--20.

\bibitem[\protect\citeauthoryear{Nielsen and Chuang}{Nielsen and Chuang}{2010}]{nielsen2010quantum}
Nielsen, M.~A. and I.~L. Chuang (2010).
\newblock {\em Quantum computation and quantum information}.
\newblock CUP.

\bibitem[\protect\citeauthoryear{O’Malley}{O’Malley}{2018}]{o2018approach}
O’Malley, D. (2018).
\newblock An approach to quantum-computational hydrologic inverse analysis.
\newblock {\em Scientific reports\/}~{\em 8\/}(1), 6919.

\bibitem[\protect\citeauthoryear{Pellow-Jarman, Sinayskiy, Pillay, and Petruccione}{Pellow-Jarman et~al.}{2021}]{pellow2021comparison}
Pellow-Jarman, A., I.~Sinayskiy, A.~Pillay, and F.~Petruccione (2021).
\newblock A comparison of various classical optimizers for a variational quantum linear solver.
\newblock {\em Quantum Information Processing\/}~{\em 20\/}(6), 202.

\bibitem[\protect\citeauthoryear{Preskill}{Preskill}{2018}]{preskill2018quantum}
Preskill, J. (2018).
\newblock Quantum computing in the nisq era and beyond.
\newblock {\em Quantum\/}~{\em 2}, 79.

\bibitem[\protect\citeauthoryear{Press}{Press}{1968}]{Press_1968}
Press, F. (1968).
\newblock {Earth models obtained by Monte-Carlo inversion}.
\newblock {\em J. Geophys. Res.\/}~{\em 73}, 5223--5234.

\bibitem[\protect\citeauthoryear{Sanders}{Sanders}{2021}]{sanders_quantum_2021}
Sanders, B.~C. (2021).
\newblock Quantum leap for quantum primacy.
\newblock {\em Physics\/}~{\em 14}, 147.

\bibitem[\protect\citeauthoryear{Schuld, Sinayskiy, and Petruccione}{Schuld et~al.}{2016}]{schuld2016prediction}
Schuld, M., I.~Sinayskiy, and F.~Petruccione (2016).
\newblock Prediction by linear regression on a quantum computer.
\newblock {\em Physical Review A\/}~{\em 94\/}(2), 022342.

\bibitem[\protect\citeauthoryear{Sevilla and Riedel}{Sevilla and Riedel}{2020}]{sevilla2020forecasting}
Sevilla, J. and C.~J. Riedel (2020).
\newblock Forecasting timelines of quantum computing.
\newblock {\em arXiv preprint arXiv:2009.05045\/}.

\bibitem[\protect\citeauthoryear{Shampine}{Shampine}{1986}]{shampine1986some}
Shampine, L.~F. (1986).
\newblock Some practical runge-kutta formulas.
\newblock {\em Mathematics of Computation\/}~{\em 46\/}(173), 135--150.

\bibitem[\protect\citeauthoryear{Shor}{Shor}{1994}]{shor1994algorithms}
Shor, P.~W. (1994).
\newblock Algorithms for quantum computation: Discrete logarithms and factoring.
\newblock In {\em Proceedings: 35th Annual Symposium on Foundations of Computer Science}, pp.\  124--134. IEEG.

\bibitem[\protect\citeauthoryear{Shor}{Shor}{1999}]{shor_polynomial-time_1997}
Shor, P.~W. (1999).
\newblock Polynomial-time algorithms for prime factorization and discrete logarithms on a quantum computer.
\newblock {\em SIAM review\/}~{\em 41\/}(2), 303--332.

\bibitem[\protect\citeauthoryear{Sivak, Eickbusch, Royer, Singh, Tsioutsios, Ganjam, Miano, Brock, Ding, and Frunzio}{Sivak et~al.}{2023}]{sivak2023real}
Sivak, V., A.~Eickbusch, B.~Royer, S.~Singh, I.~Tsioutsios, S.~Ganjam, A.~Miano, B.~Brock, A.~Ding, and L.~Frunzio (2023).
\newblock Real-time quantum error correction beyond break-even.
\newblock {\em Nature\/}~{\em 616\/}(7955), 50--55.

\bibitem[\protect\citeauthoryear{Smolin, Gambetta, and Smith}{Smolin et~al.}{2012}]{smolin2012efficient}
Smolin, J.~A., J.~M. Gambetta, and G.~Smith (2012).
\newblock Efficient method for computing the maximum-likelihood quantum state from measurements with additive gaussian noise.
\newblock {\em Physical review letters\/}~{\em 108\/}(7), 070502.

\bibitem[\protect\citeauthoryear{Souza, Martins, Roditi, S{\'a}, Sarthour, and Oliveira}{Souza et~al.}{2022}]{souza2022application}
Souza, A.~M., E.~O. Martins, I.~Roditi, N.~S{\'a}, R.~S. Sarthour, and I.~S. Oliveira (2022).
\newblock An application of quantum annealing computing to seismic inversion.
\newblock {\em Frontiers in Physics\/}~{\em 9}, 748285.

\bibitem[\protect\citeauthoryear{Suau, Staffelbach, and Calandra}{Suau et~al.}{2021}]{suau2021practical}
Suau, A., G.~Staffelbach, and H.~Calandra (2021).
\newblock Practical quantum computing: Solving the wave equation using a quantum approach.
\newblock {\em ACM Transactions on Quantum Computing\/}~{\em 2\/}(1), 1--35.

\bibitem[\protect\citeauthoryear{S{\"u}sstrunk and Huber}{S{\"u}sstrunk and Huber}{2016}]{susstrunk2016classification}
S{\"u}sstrunk, R. and S.~D. Huber (2016).
\newblock Classification of topological phonons in linear mechanical metamaterials.
\newblock {\em Proceedings of the National Academy of Sciences\/}~{\em 113\/}(33), E4767--E4775.

\bibitem[\protect\citeauthoryear{Suzuki}{Suzuki}{1976}]{suzuki_generalized_1976}
Suzuki, M. (1976).
\newblock Generalized trotter's formula and systematic approximants of exponential operators and inner derivations with applications to many-body problems.
\newblock {\em Communications in Mathematical Physics\/}~{\em 51\/}(2), 183--190.

\bibitem[\protect\citeauthoryear{Suzuki}{Suzuki}{1991}]{suzuki_general_1991}
Suzuki, M. (1991).
\newblock General theory of fractal path integrals with applications to many-body theories and statistical physics.
\newblock {\em Journal of Mathematical Physics\/}~{\em 32\/}(2), 400--407.

\bibitem[\protect\citeauthoryear{Trahan, Loveland, Davis, and Ellison}{Trahan et~al.}{2023}]{trahan2023variational}
Trahan, C.~J., M.~Loveland, N.~Davis, and E.~Ellison (2023).
\newblock A variational quantum linear solver application to discrete finite-element methods.
\newblock {\em Entropy\/}~{\em 25\/}(4), 580.

\bibitem[\protect\citeauthoryear{Trotter}{Trotter}{1959}]{trotter_product_1959}
Trotter, H.~F. (1959).
\newblock On the product of semi-groups of operators.
\newblock {\em Proceedings of the American Mathematical Society\/}~{\em 10\/}(4), 545--551.

\bibitem[\protect\citeauthoryear{Van~Dam, Mosca, and Vazirani}{Van~Dam et~al.}{2001}]{van2001powerful}
Van~Dam, W., M.~Mosca, and U.~Vazirani (2001).
\newblock How powerful is adiabatic quantum computation?
\newblock In {\em Proceedings 42nd IEEE symposium on foundations of computer science}, pp.\  279--287. IEEE.

\bibitem[\protect\citeauthoryear{Villanueva, Najafi, and Kappen}{Villanueva et~al.}{2023}]{villanueva2023adiabatic}
Villanueva, A., P.~Najafi, and H.~J. Kappen (2023).
\newblock Why adiabatic quantum annealing is unlikely to yield speed-up.
\newblock {\em Journal of Physics A: Mathematical and Theoretical\/}~{\em 56\/}(46), 465304.

\bibitem[\protect\citeauthoryear{Virieux and Operto}{Virieux and Operto}{2009}]{Virieux_2009}
Virieux, J. and S.~Operto (2009).
\newblock An overview of full waveform inversion in exploration geophysics.
\newblock {\em Geophysics\/}~{\em 74}, WCC127--WCC152.

\bibitem[\protect\citeauthoryear{Wang, Fontana, Cerezo, Sharma, Sone, Cincio, and Coles}{Wang et~al.}{2021}]{wang2021noise}
Wang, S., E.~Fontana, M.~Cerezo, K.~Sharma, A.~Sone, L.~Cincio, and P.~J. Coles (2021).
\newblock Noise-induced barren plateaus in variational quantum algorithms.
\newblock {\em Nature communications\/}~{\em 12\/}(1), 6961.

\bibitem[\protect\citeauthoryear{Watrous}{Watrous}{2018}]{watrous_2018}
Watrous, J. (2018).
\newblock {\em The theory of quantum information}.
\newblock CUP.

\bibitem[\protect\citeauthoryear{Wiebe, Braun, and Lloyd}{Wiebe et~al.}{2012}]{wiebe2012quantum}
Wiebe, N., D.~Braun, and S.~Lloyd (2012).
\newblock Quantum algorithm for data fitting.
\newblock {\em Physical review letters\/}~{\em 109\/}(5), 050505.

\bibitem[\protect\citeauthoryear{Yi and Crosson}{Yi and Crosson}{2022}]{yi_spectral_2022}
Yi, C. and E.~Crosson (2022).
\newblock Spectral analysis of product formulas for quantum simulation.
\newblock {\em npj Quantum Information\/}~{\em 8\/}(1), 1--6.

\end{thebibliography}

\appendix

\section{Continuous transformation}\label{A:Continuous transformation}

Here, we construct an invertible transformation $\mathcal{T}$ that transforms the continuous 1-D elastic wave equation into a continuous Schrödinger equation. We consider functions $a(x)$ that are differentiable at least once and satisfy either $a(0)=f(W) = 0$, $a(0)=0$ and $\partial_x a(W)=0$, or  $\partial_x a(0)=0$ and $a(W)=0$, which define three different Hilbert spaces $\mathbb{H}_{D,D}$, $\mathbb{H}_{D,N}$ or $\mathbb{H}_{N,D}$ referring to either two Dirichlet or at least one Dirichlet boundary condition. All three Hilbert spaces are equipped with the natural inner product
\begin{equation}
    \braket{a,b} = \int_0^Wdx\, a^*(x)b(x), \quad a,b \in \mathbb{H}_{B,B'}\,,
\end{equation}
for $B, B' \in \{D, N\}$, and where $*$ denotes complex conjugation. We introduce the adjoint operator, $\mathcal{A}^{\dagger}$ of some operator $\mathcal{A}$, corresponding to the above inner product and satisfying
\begin{equation}
\braket{a,\mathcal{A}b} =   \braket{\mathcal{A}^{\dagger}a,b} , \quad a,b \in \mathbb{H}_{B,B'}\,.
\end{equation}
An operator, $\mathcal{A}$ is self-adjoint, i.e. $\mathcal{A}^{\dagger}=\mathcal{A}$, if $\braket{a,\mathcal{A}b} =\braket{\mathcal{A}a,b}$, 
and anti-self-adjoint if $\braket{a,\mathcal{A}b} =-\braket{\mathcal{A}a,b}$, i.e. $\mathcal{A}^{\dagger}=-\mathcal{A}$. For two operators $\mathcal{A}$ and $B$ we have that $(\mathcal{A}\mathcal{B})^{\dagger} = \mathcal{B}^{\dagger}\mathcal{A}^{\dagger}$. Furthermore, in $\mathbb{H}_{D, D}$, $\mathbb{H}_{D, N}$ and $\mathbb{H}_{N, D}$ we have $\partial_x^{\dagger} = -\partial_x$, which can be established through integration by parts and invoking the boundary conditions. With these properties of the adjoint operator we can express the mass-transformed wave operator as 
\begin{equation}\label{eq: density transformed waveop 2}
     \mathcal{L}_{\rho}:=\rho^{-1/2}\partial_x[\mu \partial_x]\rho^{-1/2} = -\big[\mu^{1/2}\partial_x\rho^{-1/2}\big]^{\dagger}\mu^{1/2}\partial_x\rho^{-1/2}\,.
\end{equation}
The differentiation operator $\partial_x$ has inverses given by the indefinite integral $\int dx: \, f(x) \rightarrow F(x) + C$, with a constant $C$ and $\partial_xF(x) = f(x)$. Assuming at least one Dirichlet boundary, which results in the Hilbert spaces considered here, any wave field has zero offset $C=0$, rendering the inverse unique. We can, therefore, construct 
\begin{equation}
\mathcal{T} := \begin{bmatrix}
  \mu^{1/2}\partial_x\rho^{-1/2} & 0\\
  0 & 1
\end{bmatrix}\,,
\end{equation}
with its right inverse given by
\begin{equation}
\mathcal{T}^{-1} = \begin{bmatrix}
  \rho^{1/2}\int dx \, \mu^{-1/2} & 0\\
  0 & 1
\end{bmatrix}\,.
\end{equation}
This leads to 
\begin{equation}
    \mathcal{T}\mathcal{Q}\mathcal{T}^{-1} =
    \mathcal{T}\begin{bmatrix}
      0 & 1\\
      \mathcal{L}_{\rho} & 0
   \end{bmatrix}\mathcal{T}^{-1}
    =\begin{bmatrix}
  0 & \mu^{1/2}\partial_x\rho^{-1/2}\\
  -\big[\mu^{1/2}\partial_x\rho^{-1/2}\big]^{\dagger} & 0
\end{bmatrix}
= \begin{bmatrix}
  0 & \mu^{1/2}\partial_x\rho^{-1/2}\\
  \rho^{1/2}\partial_x\mu^{-1/2} & 0
\end{bmatrix}\,,
\end{equation}
which is clearly anti-self-adjoint under the quantum inner product \eqref{eq: inner product augm} of the quantum Hilbert space, $\mathbb{H}^q = \mathbb{H}_{B,B'}\oplus\mathbb{H}_{B,B'}$. Finally, we conclude that $H = i\mathcal{T}\mathcal{Q}\mathcal{T}^{-1}$ is therefore self-adjoint.

\section{Complexity of sparse Hamiltonian simulation}\label{A:Sparse Hamiltonian simulation}

The first step of the stated HS approach is setting up the initial state according to \eqref{eq: prepare initial state}. Today's best sparse state initialisation methods need $\mathcal{O}(nS^2\log(S))$ classical operations,
producing a circuit with $\mathcal{O}(nS)$ quantum gates,
where $S$ is the number of non-zero initial conditions \citep{gleinig2021efficient}.
This cost can be amortized with relative ease if $S \ll 2^n$,
which matches typical use cases. 

To provide insight into the complexity of the time evolution itself using the currently best available methods, we first need to discuss the implementation of sparse matrices. It is generally assumed that sparse matrices can efficiently be accessed using quantum oracles. A quantum oracle is a function solving a subroutine of an algorithm that has an efficient implementation, meaning it has a polynomial runtime complexity and hence its circuit depth $D$ is in $\mathcal{O}(n^k)$. The existence of these oracles is generally assumed in quantum algorithmic research and allows for the analysis and improvement of subroutines in a larger algorithmic framework. A quantum oracle must be a unitary operation, and it can be shown that this requirement can be satisfied for any classical function by augmenting both its input and output with additional helper qubits appropriately. To represent a $d$-sparse matrix $\mathbf{X}$, two oracles are used: (1) one that takes a row index $i$ and column index $j$, and returns the value $\mathbf{X}(i, j)$, and (2) one that takes a row index $i$ and nonzero index $l \in [0, d)$, and returns the column index $j$ of the $l$-th nonzero element in the row $i$.

As described in section \ref{S:Hamiltonian simulation},
the action of the time-evolution operator $e^{-i\mathbf{H}t}$ is carried out only approximately because the exact matrix exponential evaluation is prohibitively expensive.
There are multiple algorithms available that realize this approximation, the most efficient so far being HS by qubitization \citep{low2019hamiltonian}.
The input $d$-sparse Hamiltonian $\mathbf{H}$ can be provided using the two oracles from above.
For the time $t$ and error $\varepsilon$, the qubitization algorithm needs $\mathcal{O}(dt ||\mathbf{H}||_\mathrm{max} + \log(1/\varepsilon))$ queries to the oracles.
Here, $||\mathbf{H}||_\mathrm{max}$ is the maximum absolute value of $\mathbf{H}$. 
The error $\varepsilon$ is the desired fidelity of the time-evolution operator, more specifically, the maximum admissible $||\mathbf{U}' - \mathbf{U}||_2$, where $\mathbf{U} = e^{-it\mathbf{H}}$ denotes the exact time-evolution operator, and $\mathbf{U}'$ denotes its approximation produced by the algorithm.
While HS is an area of active research, the query complexity of this algorithm already reaches fundamental lower bounds in the $\mathcal{O}$-notation.
In the presented 1-D case, $d=2$, as we have a 2-sparse Hamiltonian, and
$$
||\mathbf{H}||_\mathrm{max}
= ||\mathbf{U}||_\mathrm{max}
= ||\mathbf{E}^{1/2}\mathbf{D}\mathbf{M}^{-1/2}||_\mathrm{max}
= \max_{i=0}^{2^n-1}{\sqrt{\frac{\mu(x_i)}{\rho(x_i)}}}.
$$
Hence, the query complexity is linear in $d$, $t$, $||\mathbf{H}||_\mathrm{max}$ and $1/\varepsilon$.
Assuming that $\varepsilon$, $d$ and $\mathbf{H}_\textrm{max}$ do not depend on $n$,
which clearly holds in our case, the complexity of the overall algorithm is polynomial in $n$ and logarithmic in $N$. This establishes the exponential speedup over classical solutions.

A more technical, in-depth analysis of the sparse HS utilized for simulating coupled classical oscillators
and proofs establishing exponential speedup against classical algorithms can be found in \cite{babbush2023exponential} and a general complexity analysis for simulating $d$-sparse Hamiltonians in \cite{low2019hamiltonian}.

\section{General Boundary Conditions}\label{A:General Boundary Conditions}

The derived Hamiltonian formulation based on the FD matrix \eqref{eq: FD matrix} naturally results in Dirichlet and Neumann boundary conditions (BCs) on the right and left side of the domain, respectively. We now extend this approach to arbitrary BCs, with at least one Dirichlet BC to ensure the negative definiteness of the stiffness matrix. To that end, we define the non-square first-order-accurate forward finite-difference matrix $\mathbf{D} $ as
\begin{equation} \label{eq: FD matrix GB}
   \mathbf{D} = \frac{1}{\Delta x} \begin{bmatrix}
  d_\textrm{left} & 0 & 0 & \dots & 0 \\
  1 & -1 & 0 & \dots & 0 \\
  0 & 1 & -1 & \dots & 0 \\
  0 & 0 & 1 & \dots & 0 \\
  \vdots & \vdots & \vdots & \ddots & \vdots \\
  0 & 0 & 0 & \dots 1 & -1 \\
  0 & 0 & 0 & \dots 0 & d_\textrm{right}
  \end{bmatrix} \in \mathbb{R}^{2^{n-1}\times(2^{n-1}-1)}.
\end{equation}
For Dirichlet-Dirichlet boundary conditions, $d_\textrm{left} = d_\textrm{right} = 1$. If one of the BCs is Neumann, then we set respectively $d_\textrm{left} = 0$ or $d_\textrm{right} = 0$. The symmetric negative-definite stiffness matrix $\mathbf{K}$ is obtained as
\begin{equation}
    \mathbf{K} = -\mathbf{D}^T\mathbf{E} \mathbf{D} \in \mathbb{R}^{(2^{n-1}-1)\times(2^{n-1}-1)},
\end{equation}
where $\mathbf{E} \in \mathbb{R}^{2^{n-1}\times2^{n-1}}$. We obtain the mass-transformed stiffness matrix $\tilde{\mathbf{K}}$ as
\begin{equation}
    \tilde{\mathbf{K}} = \mathbf{M}^{-1/2} \mathbf{K} \mathbf{M}^{-1/2} = -\mathbf{U}^T \mathbf{U} \in \mathbb{R}^{(2^{n-1}-1)\times(2^{n-1}-1)}
\end{equation}
with $\mathbf{M}^{-1/2} \in \mathbb{R}^{(2^{n-1}-1)\times(2^{n-1}-1)}$, and consequently 
\begin{equation}
    \mathbf{U} = \mathbf{E}^{1/2}\mathbf{D}\mathbf{M}^{-1/2} \in \mathbb{R}^{2^{n-1}\times(2^{n-1}-1)}.
\end{equation}
We can now define the transformation
\begin{equation}
    \mathbf{T} = \begin{bmatrix}\mathbf{U} & \mathbf{0}\\ \mathbf{0} & \mathbf{\mathds{1}}_{2^{n-1}\times(2^{n-1}-1)} \end{bmatrix} \in \mathbb{R}^{2^{n}\times(2^{n}-2)},
\end{equation}
which has full column rank and hence a unique left inverse $\mathbf{T}^{+}$,
i.e., $\mathbf{T}^{+}\mathbf{T}=\mathbf{I}$. 
With the impedance matrix $\mathbf{Q}$ defined as
\begin{equation}
    \mathbf{Q} = \begin{bmatrix} \mathbf{0} & \mathbf{\mathds{1}}_{(2^{n-1}-1)\times(2^{n-1}-1)}\\ \tilde{\mathbf{K}} & \mathbf{0} \end{bmatrix} \in \mathbb{R}^{(2^{n}-2)\times(2^{n}-2)},
\end{equation}
we finally obtain the Hermitian Hamiltonian $\mathbf{H} \in \mathbb{R}^{2^n\times2^n}$ as
\begin{equation}
    \mathbf{H} = i\mathbf{T}\mathbf{Q}\mathbf{T}^{+} = i\begin{bmatrix}\mathbf{U} & \mathbf{0}\\ \mathbf{0} & \mathbf{\mathds{1}}_{2^{n-1}\times(2^{n-1}-1)} \end{bmatrix} \begin{bmatrix} \mathbf{0} & \mathbf{\mathds{1}}_{(2^{n-1}-1)\times(2^{n-1}-1)}\\ -\mathbf{U}^T \mathbf{U} & \mathbf{0} \end{bmatrix} \begin{bmatrix}\mathbf{U}^+ & \mathbf{0}\\ \mathbf{0} & \mathbf{\mathds{1}}_{(2^{n-1}-1)\times2^{n-1}} \end{bmatrix},
\end{equation}
which we use in the HS. Note this formulation allows for $2^{n-1}-1$ grid points for $n$ qubits.

\section{Quantum Circuit Implementation}\label{A:Quantum Circuit Implementation}

\begin{figure}
    \centering
    \begin{subfigure}[b]{0.45\textwidth}
        \centering
        \includegraphics[width=2.7\linewidth, angle=-90]{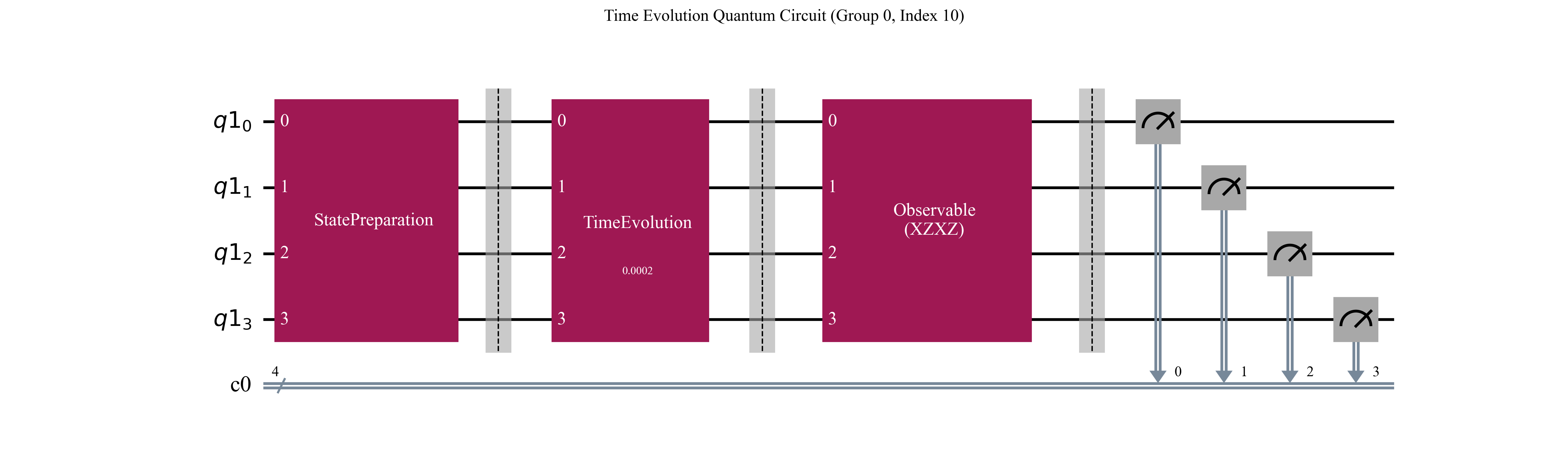}
        \caption{Simplified Circuit.}
        \label{fig:circuit}
    \end{subfigure}
    \hfill
    \begin{subfigure}[b]{0.45\textwidth}
        \centering
        \includegraphics[width=2.1\linewidth, angle=-90]{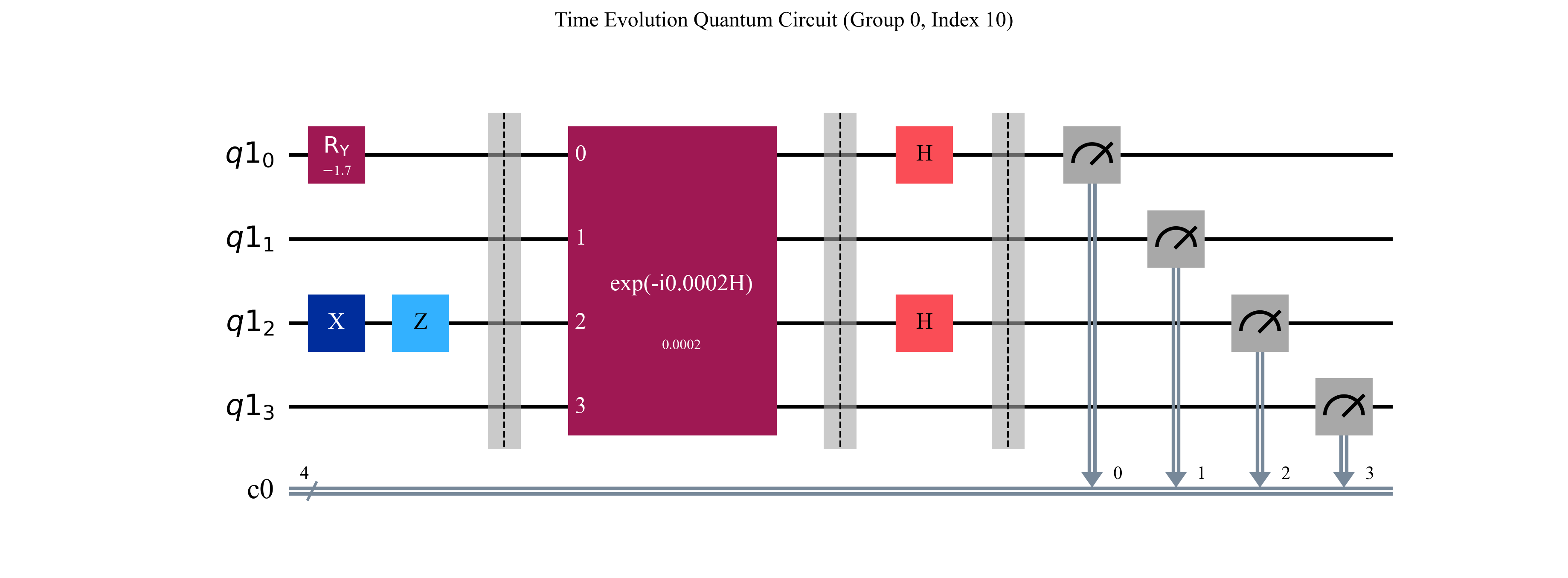}
        \caption{Expanded Circuit}
        \label{fig:circuit_p}
    \end{subfigure}
    \caption{Circuit Representation of the wave simulation algorithm for a 7 grid point problem subject to the XZXZ Pauli observable. The circuits show four distinct 
    algorithmic stages in differing levels of detail: 1.) The state preparation of $\ket{\psi(0)}$, 2.) the time evolution from $\ket{\psi(0)} \rightarrow \ket{\psi(t)}$, 3.) the application of the observable rotating the state into the computational basis, and 4.) a full state measurement of all qubits in the computational basis.}
    \label{fig:two_circuits}
\end{figure}

Figure \ref{fig:two_circuits} displays the quantum circuit implemented in section \ref{S:Implementation} in the standard circuit model notation for one randomly chosen observable. While Figure \ref{fig:circuit} shows the circuit in a general and simplified form, Figure \ref{fig:circuit_p} shows the detailed circuit implementation for the circuit run on the QC IBM Brisbane. In both circuit models, we can observe the four distinct subroutines included in the full quantum algorithm: 1.) The state preparation of $\ket{\psi(0)}$, 2.) the time evolution from $\ket{\psi(0)} \rightarrow \ket{\psi(t)}$, 3.) the application of the observable to bring the state into the computational basis, and 4.) a full state measurement of all qubits in the computational basis.

After initializing the qubits in state $\ket{0}$, a state preparation routine is employed. This sub-circuit prepares the initial wave field state from  $\ket{0} \rightarrow \ket{\psi(0)}$ (\ref{eq: prepare initial state}). Generally, this state initialization can be done using sparse initialization routines (\cite{gleinig2021efficient}), or simplified routines to initialize highly symmetric states with minimal circuit overhead. A variant of a simplified initialization routine is displayed in Figure \ref{fig:circuit_p}. By using only three additional gates, it allows the initialization of a point-localized initial condition in the center of the simulated domain. We employ this simplified initialization routine in the simulation used for obtaining Figure \ref{fig: fordward_sim_b}.

As a next step, the time evolution circuit is implemented. This circuit evolves the state $\ket{\psi(0)}$ to $\ket{\psi(t)}$ (\ref{eq: time evolution}). We construct the time evolution circuit from the Hamiltonian $H$ and the evolution time $t$ using a scaling and squaring method (\cite{almohy2010scaling}). We do not provide a detailed circuit representation of the time evolution sub-circuit as even for small problem sizes it contains hundreds of gates arranged in a non-trivial layout.

Finally, we implement an observable according to the QST requirements. As we need to evaluate $\ket{\psi(t)}$ with respect to $2^n$ quantum observables for every time step (see section \ref{S:Implementation}), the circuits displayed in Figure \ref{fig:two_circuits} describe only one of the many necessary observable sub-circuit layouts evaluated on the QC. As each circuit evaluation only gives one measurement sample, every circuit layout needs to be evaluated multiple times to obtain statistically accurate results. To implement an observable given as a Pauli matrix on a QC it is necessary to rotate all qubits into the computational basis or $Z$-basis. The rotation directions are defined by the respective Pauli matrix components for each qubit. Figure \ref{fig:circuit_p} shows an implementation of the Pauli observable $XZXZ$, which equates to implementing Hadamard gates ($H$) for every $X$ component, and no additional gates for every $Z$ component. This sub-circuit achieves a rotation of all qubits into the $Z$-basis, where they can be measured.

As a last step, all qubits are measured and the results are subsequently used in the QST process.

The full circuit implementation is available at \url{https://github.com/malteschade/Quantum-Wave-Equation-Solver}.

\FloatBarrier

\end{document}